\documentclass[nonacm, sigconf]{acmart}

\AtBeginDocument{%
  \providecommand\BibTeX{{%
    \normalfont B\kern-0.5em{\scshape i\kern-0.25em b}\kern-0.8em\TeX}}}

\setcopyright{acmcopyright}

\usepackage{subfigure}
\usepackage{pgfplots}

\usepackage{pifont}

\usepackage{caption}

\definecolor{dark-green}{rgb}{0,0.6,0}

\definecolor{amaranth}{rgb}{0.9, 0.17, 0.31}
\definecolor{bostonuniversityred}{rgb}{0.8, 0.0, 0.0}
\definecolor{brightpink}{rgb}{1.0, 0.0, 0.5}
\definecolor{darklava}{rgb}{0.28, 0.24, 0.2}
\definecolor{darkgreen}{rgb}{0.0, 0.2, 0.13}
\definecolor{coolblack}{rgb}{0.0, 0.18, 0.39}
\definecolor{blue-violet}{rgb}{0.54, 0.17, 0.89}

\copyrightyear{2022}
\acmYear{2022}
\setcopyright{acmcopyright}\acmConference[UIST ’22]{35th Annual ACM Symposium on User Interface Software and Technology (UIST ’22)}{October 29 - November 2, 2022}{Bend, Oregon, USA}
\acmBooktitle{35th Annual ACM Symposium on User Interface Software and Technology (UIST ’22), October 29 - November 2, 2022, Bend, Oregon, USA}
\acmPrice{15.00}
\acmDOI{10.1145/3411763.3451632}
\acmISBN{978-1-4503-8095-9/21/05}

\newcommand{\grease}{\textit{GreaseVision}}

\begin{document}

\title[\grease{}]{\grease{}: Rewriting the Rules of the Interface}

\author{Siddhartha Datta}
\email{siddhartha.datta@cs.ox.ac.uk}
\affiliation{%
  \institution{University of Oxford}
}

\author{Konrad Kollnig}
\email{konrad.kollnig@cs.ox.ac.uk}
\affiliation{%
  \institution{University of Oxford}
}

\author{Nigel Shadbolt}
\email{nigel.shadbolt@cs.ox.ac.uk}
\affiliation{%
  \institution{University of Oxford}
}

\begin{abstract}
%
Digital harms can manifest across any interface.
Key problems in addressing these harms include the high individuality of harms and the fast-changing nature of digital systems.
As a result, we still lack a systematic approach to 
study harms and produce interventions for end-users.
%
We put forward \grease{}, 
a new framework that enables end-users to collaboratively develop interventions against harms in software using a no-code approach and recent advances in few-shot machine learning.
The contribution of the framework and tool 
allow individual end-users to study their usage history and create personalized interventions.
Our contribution also enables researchers to study the distribution of harms and interventions at scale.
\end{abstract}



\maketitle

\section{Introduction}
\label{sec:intro}

The design of good user interfaces can be challenging. One usually employs a range of qualitative methods, including interviews, surveys and user stories, paired with quantitative insights from analytics tools to understand user needs.
In a fast changing world, however, with sometimes highly individual needs, traditional one-fits-all software development faces difficulty in keeping up with the pace of change and the breadth of user requirements.
At the same time, the digital world is rife with a range of harms, ranging from dark patterns to hate speech and violence. Current mitigation strategies predominantly use legal tools to target the providers of social media platforms and other websites to combat these threats.
While potentially potent, the use of legal remedies often struggles to capture the language of computer programs, which are written in \textit{code}, and not in \textit{law}.

This paper is taking a step back to improve the user experience in the digital world.
To achieve this, we put forward a new design philosophy for the development of software interfaces that serves its users. We call our technical prototype \textit{GreaseVision}.
Specifically, with our approach, we allow users to change problems about the software that they use \textit{by themselves}.
No prior programming knowledge is required; instead, we leverage recent advances in low-code programming to allow individuals avoid harmful UI aspects.
As such, our work introduces a new way to fix flaws in code that does not rely on legal mechanisms, can be applied quickly, and can keep up with the fast-changing development cycle in the digital world.

For the research community in human-computer interaction, our approach presents a new valuable resource for studying and improving interface design. Users can develop by themselves, but not necessarily just for themselves. Instead, users are able to share their interventions easily with others, thereby creating an online repository of widespread harms in apps and effective interventions against them.
If one gathers all the different interventions that users develop for popular websites such as Facebook or Twitter, this would yield novel insights into the problems with current interface design.



\noindent\textbf{Contributions: }
Our work aims to contribute a novel interface modification framework, which we call \grease{}.
At a structural-level, 
our framework enables end-users to develop personalized interface modifications, either individually or collaboratively.
This is supported by the use of screenome visualization, human-in-the-loop learning, and an overlay/hooks-enabled low-code development platform.
Within the defined scopes, 
we enable the aggregation of distributionally-wide end-user digital harms (self-reflection for end-users, or analyzing the harms repository for researchers),
to further enable the modification of user interfaces across a wide range of software systems, supported by the usage of visual overlays, autonomously developed by users, and enhanced by scalable machine learning techniques.
We publicly provide complete and reproducible implementation details to enable researchers to not only study harms and interventions, but other interface modification use cases as well.

\noindent\textbf{Structure: }
Having introduced the challenge of end-user interface modification in Section~\ref{sec:intro}, 
we detail the landscape of problems and opportunities manifesting in digital harms and their interface modification methods in Section~\ref{sec:background}.
We set the scope for the paper in Section~\ref{sec:scope}, 
and share our proposed method~--~\grease{}~--~in Section~\ref{sec:grease}.
We evaluate and discuss our method and findings in Section~\ref{sec:evaluation} and~\ref{sec:discussion} respectively, 
and share final thoughts and conclusions in Section~\ref{sec:conclusion}.



\section{Background}
\label{sec:background}

\begin{figure*}
    \centering
    \caption{Architecture of \textit{GreaseTerminator} (left) and \textit{GreaseVision} (right).}
    \Description{Architecture of \textit{GreaseTerminator} (left) and \textit{GreaseVision} (right).}
    \subfigure[The high-level architecture of \textit{GreaseTerminator}. Details are explained in Section 2.3 and 4.2. 
    ]{
    \includegraphics[width=0.425\linewidth]{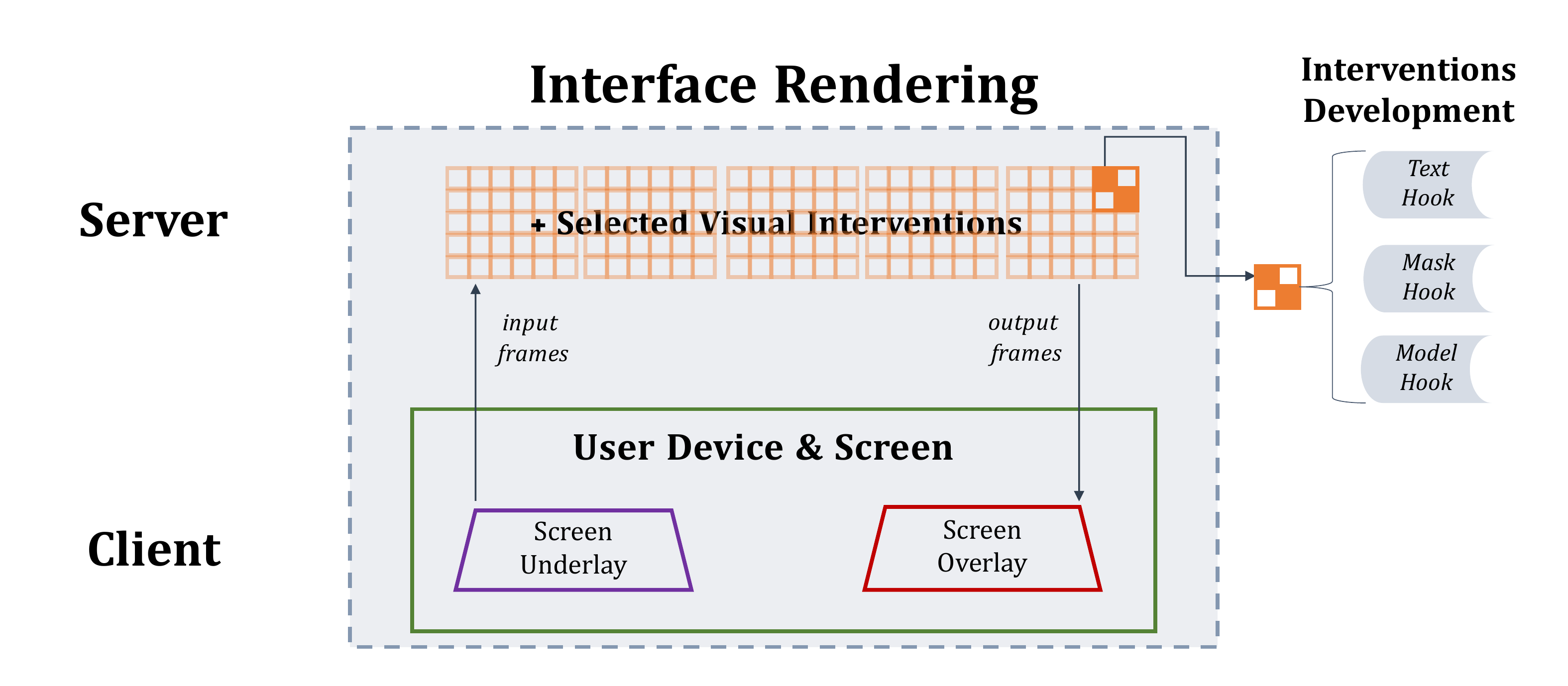}
    }
    \hfill
    \subfigure[The high-level architecture of \grease{}, both as a summary of our technical infrastructure as well as one of the collaborative HITL interventions development approach.]{
    \includegraphics[width=0.55\linewidth]{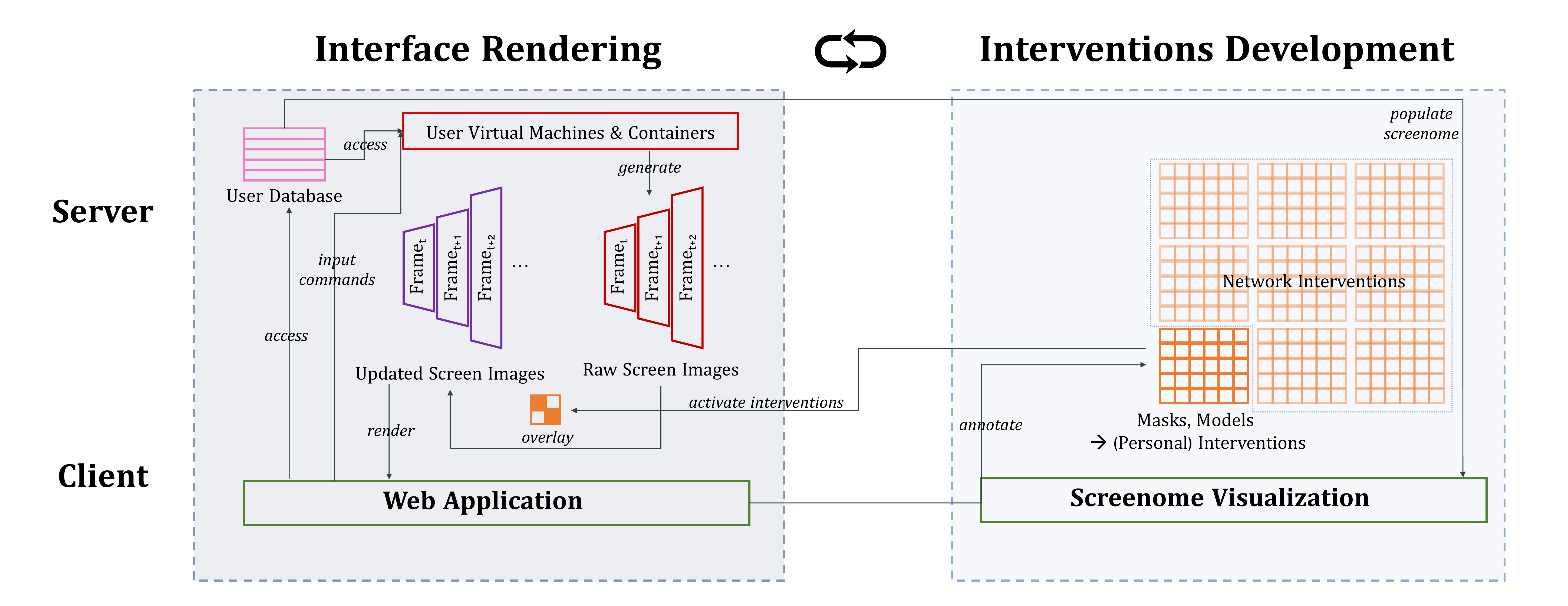}
    }
    \label{fig:archi_gw}
\end{figure*}

\subsection{Motivation: Pervasiveness and Individuality of Digital Harms}

It is well-known that digital harms are widespread in our day-to-day technologies.
Despite this, the academic literature around these harms is still developing, and it remains difficult to state exactly what the harms are that need to be addressed.
Famously, Gray et al.~\citep{10.1145/3173574.3174108} put forward a 5-class taxonomy to classify dark patterns within apps:
\textit{interface interference} (elements that manipulate the user interface to induce certain actions over other actions), 
\textit{nagging} (elements that interrupt the user’s current task with out-of-focus tasks)
\textit{forced action} (elements that introduce sub-tasks forcefully before permitting a user to complete their desired task), 
\textit{obstruction} (elements that introduce subtasks with the intention of dissuading a user from performing an operation in the desired mode),
and \textit{sneaking} (elements that conceal or delay information relevant to the user in performing a task).

A challenge with such framework and taxonomies is to capture and understand the material impacts of harms on individuals.
Harms tend to be highly individual and vary in terms of how they manifest within users of digital systems.
The harms landscape is also quickly changing with ever-changing digital systems.
Defining the spectrum of harms is still an open problem, 
the range varying from
heavily-biased content (e.g. disinformation, hate speech),
self-harm (e.g. eating disorders, self-cutting, suicide),
cyber crime (e.g. cyber-bullying, harassment,
promotion of and recruitment for extreme causes (e.g. terrorist organizations),
to demographic-specific exploitation (e.g. child-inappropriate content, social engineering attacks)~\citep{hmgov, 10.1145/2998181.2998224, 10.1145/3038912.3052555, 10.1145/3313831.3376370, 10.1145/3359186},
for which we recommend the aforementioned cited literature.
The last line of defense against many digital harms is the user interface.
This is why we are interested in interface-emergent harms in this paper, and how to support individuals in developing their own strategies to cope with and overcome such harms.

\subsection{Developments in Interface Modification \& Re-rendering}
\label{sec:existing_methods}

Digital harms have long been acknowledged as a general problem, and a range of technical interventions against digital harms are developed. 
Interventions, also similarly called modifications or patches, are changes to the software, which result in a change in (perceived) functionality and end-user usage.
We review and categorize key \textit{technical} intervention methods for interface modification by end-users, with cited examples specifically for digital harms mitigation.
While there also exist non-technical interventions, in particular legal remedies, it is beyond this work to give a full account of these different interventions against harms; a useful framework for such an analysis is provided by Lawrence Lessig~\citep{lessig_code_2006} who characterised the different regulatory forces in the digital ecosystem.

\textbf{Interface-code modifications}~\citep{gd,apkm,jeon_dr_2012,rasthofer_droidforce_2014,davis_retroskeleton_2013,garcia-alfaro_appguard_2014,Aurasium,lp,Davis12i-arm-droid:a,lyngs_i_2020, cydia,xp,agarwal_protectmyprivacy_2013,enck_taintdroid_2010,InstaPrefs, vx} 
make changes to source code,
either installation code (to modify software before installation),
or run-time code (to modify software during usage).
On desktop, this is done through \textit{browser extensions} and has given rise to a large ecosystem of such extensions.
Some of the most well-known interventions are ad blockers, and tools that improve productivity online (e.g. by removing the Facebook newsfeed~\citep{lyngs_i_2020}).
On mobile, a prominent example is \textit{AppGuard}~\citep{garcia-alfaro_appguard_2014}, a research project by Backes et al. that allowed users to improve the privacy properties of apps on their phone by making small, targeted modification to apps' source code.
Another popular mobile solution in the community is the app \textit{Lucky Patcher}~\citep{lp} that allows to get paid apps for free, by removing the code relating to payment functionality directly from the app code.
    
Some of these methods may require the highest level of privilege escalation to make modifications to the operating system and other programs/apps as a root user.
On iOS, \textit{Cydia Substrate}~\cite{cydia} is the foundation for jailbreaking and further device modification. A similar system, called \textit{Xposed Framework}~\citep{xp}, exists for Android.
To alleviate the risks and challenges afflicted with privilege escalation,
\textit{VirtualXposed}~\citep{vx}
create a virtual environment on the user's Android device with simulated privilege escalation. Users can install apps into this virtual environment and apply tools of other modification approaches that may require root access. 
\textit{ProtectMyPrivacy}~\citep{agarwal_protectmyprivacy_2013} for iOS and \textit{TaintDroid}~\citep{enck_taintdroid_2010} for Android both extend the functionality of the smartphone operating system with new functionality for the analysis of apps' privacy features.
On desktops, code modifications tend not to be centred around a common framework, but are more commonplace in general due to the traditionally more permissive security model compared to mobile.
Antivirus tools, copyright protections of games and the modding of UI components are all often implemented through interface-code modifications.

\textbf{Interface-external modifications}~\citep{kovacs_thesis,bodyguard,10.1007/s10916-013-0001-1,10.1145/2675133.2675244, 10.1145/2968219.2971591,10.1145/2858036.2858403,10.1145/2541831.2541870,al,10.1145/3229434.3229463} are the arguably most common way to change default interface behaviour.
An end-user would install a program so as to affect other programs/apps.
No change to the operating system or the targeted programs/apps is made, so an uninstall of the program providing the modification would revert the device to the original state.
This approach is widely used to track duration of device usage, send notifications to the user during usage (e.g. timers, warnings), block certain actions on the user device, and other aspects.
The \textit{HabitLab}~\citep{kovacs_thesis} is a prominent example developed by Kovacs et al. at Stanford.
This modification framework is open-source and maintained by a community of developers, and provides interventions for both desktop and mobile devices.

\textbf{Visual overlay modifications} render graphics on an overlay layer over any active interface instance, including browsers, apps/programs, videos, or any other interface in the operating system. The modifications are visual, and do not change the functionality of the target interface. 
It may render sub-interfaces, labels, or other graphics on top of the foreground app.
Prominent examples are \textit{DetoxDroid}~\citep{detoxdroid},
\textit{Gray-Switch}~\citep{grayswitch}, \textit{Google Accessibility Suite}~\citep{accessibility}, and
\textit{GreaseTerminator}~\citep{greaseterminator}.

We would like to establish early on that we pursue a \textit{visual overlay modifications} approach. 
Interventions should be rendered in the form of overlay graphics based on detected elements, rather than implementing program code changes natively, hence focused on changing the interface rather than the functionality of the software.
Interventions should be generalizable; they are not solely website- or app-oriented, but \textit{interface-oriented}. 
Interventions do not target specific apps, but general interface elements and patterns that could appear across different interface environments. 
To support the systemic requirements in Section 2.4,
we require an interface modification approach that is (i) interface-agnostic and (ii) easy-to-use.
To this extent, we build upon the work of \textbf{\textit{GreaseTerminator}} \cite{greaseterminator}, a framework optimized for these two requirements.

In response to the continued widespread presence of interface-based harms in digital systems, Datta et al.~\citep{greaseterminator} developed \textit{GreaseTerminator}, a visual overlay modification method.
This approach enables researchers to develop, deploy and study interventions against interface-based harms in apps.
This is based on the observation that it used to be difficult in the past for researchers to study the efficacy of different intervention designs against harms within mobile apps (most previous approaches focused on desktop browsers).
\textit{GreaseTerminator} provides a set of `hooks' that serve as templates for researchers to develop interventions, which are then deployed and tested with study participants.
\textit{GreaseTerminator} interventions usually come in the form of machine learning models that build on the provided hooks, automatically detect harms within the smartphone user interface at run-time, and choose appropriate interventions (e.g. a visual overlay to hide harmful content, or content warnings).
A visualisation of the \textit{GreaseTerminator} approach is shown in Figure~\ref{fig:archi_gw}(a).


In addition to the contributions stated in Section 1, 
\grease{} is an improved visual overlay modification approach with respect to interface-agnosticity and ease of use. 
We discuss the technical improvements upon \textit{GreaseTerminator} in Section 4.1, 
specifically latency, device support, and interface-agnosticity.
We discuss the specific aspects of \textit{GreaseTerminator} we adopt in \grease{} in Section 4.2, specifically inter-operable hooks/overlay mechanisms.

\begin{figure*}
    \centering
    \caption{Demonstration of hooks adapted in \grease{}, based on the \textit{GreaseTerminator} implementation. }
    \Description{Demonstration of mask and model hooks, based on the \textit{GreaseTerminator} hooks implementation and adapted in \grease{}.}
    \subfigure[Occlusion of recommended items on Twitter (before \textit{left}, after \textit{right})]{
    \includegraphics[height=4.8cm]{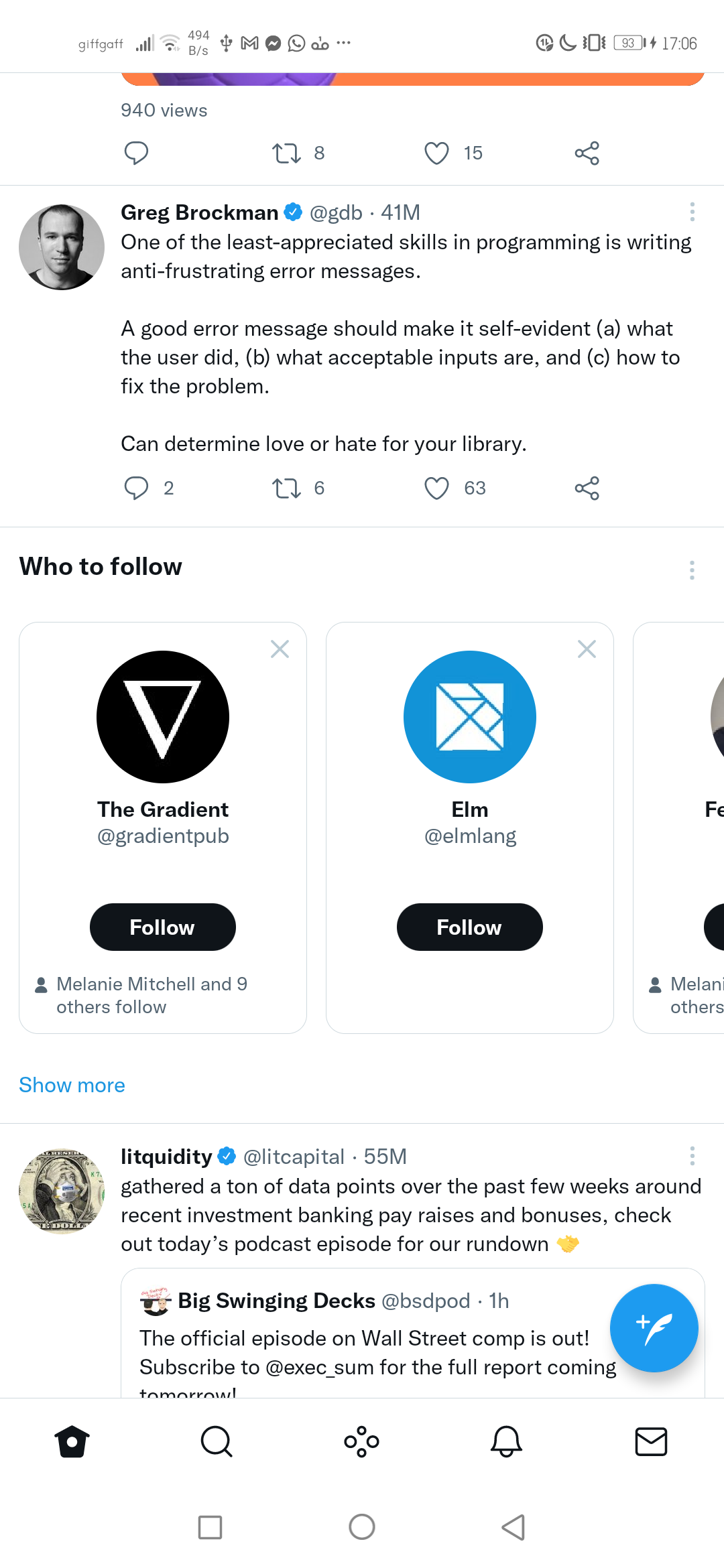}
    \includegraphics[height=4.8cm]{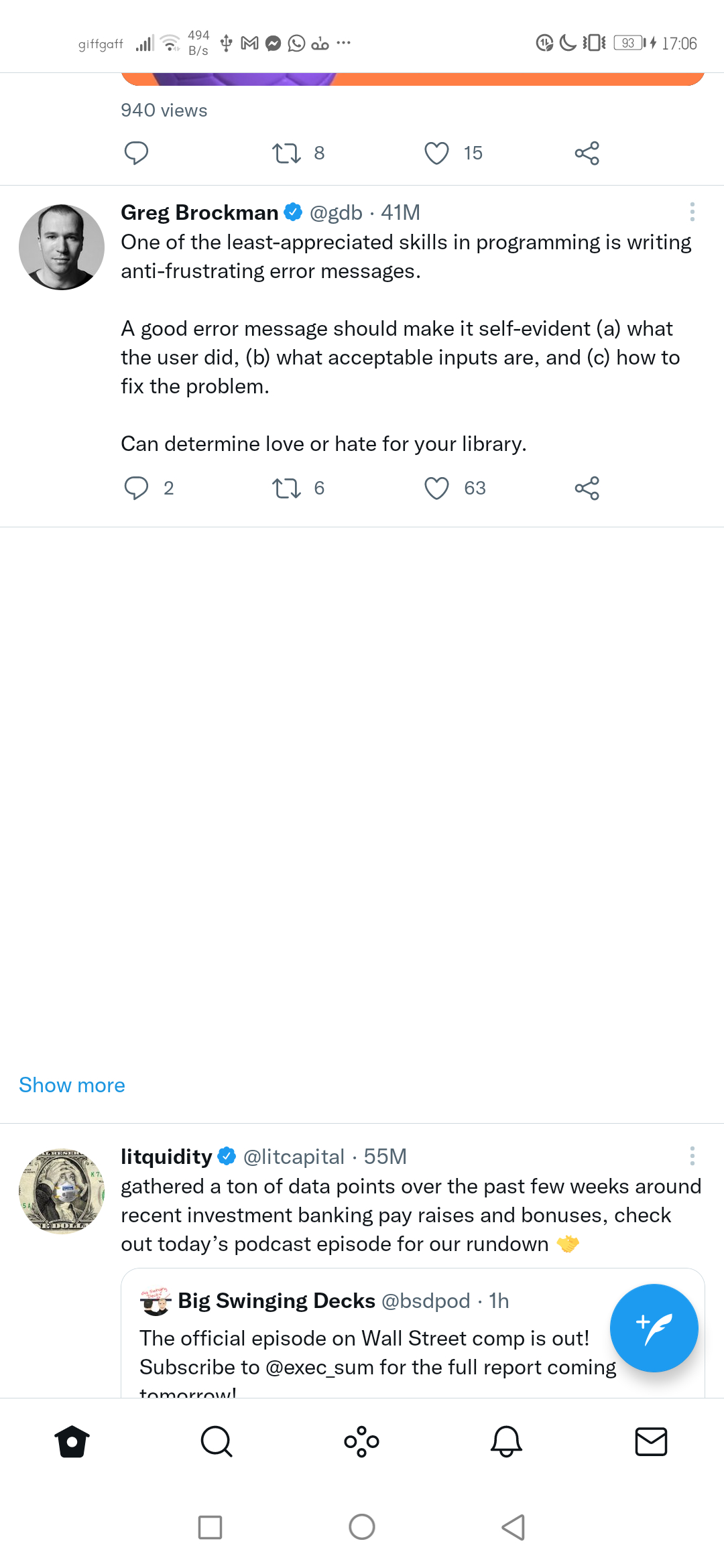}
    }
    \hfill
    \subfigure[Occlusion of recommended items on Instagram (before \textit{left}, after \textit{right})]{
    \includegraphics[height=4.8cm]{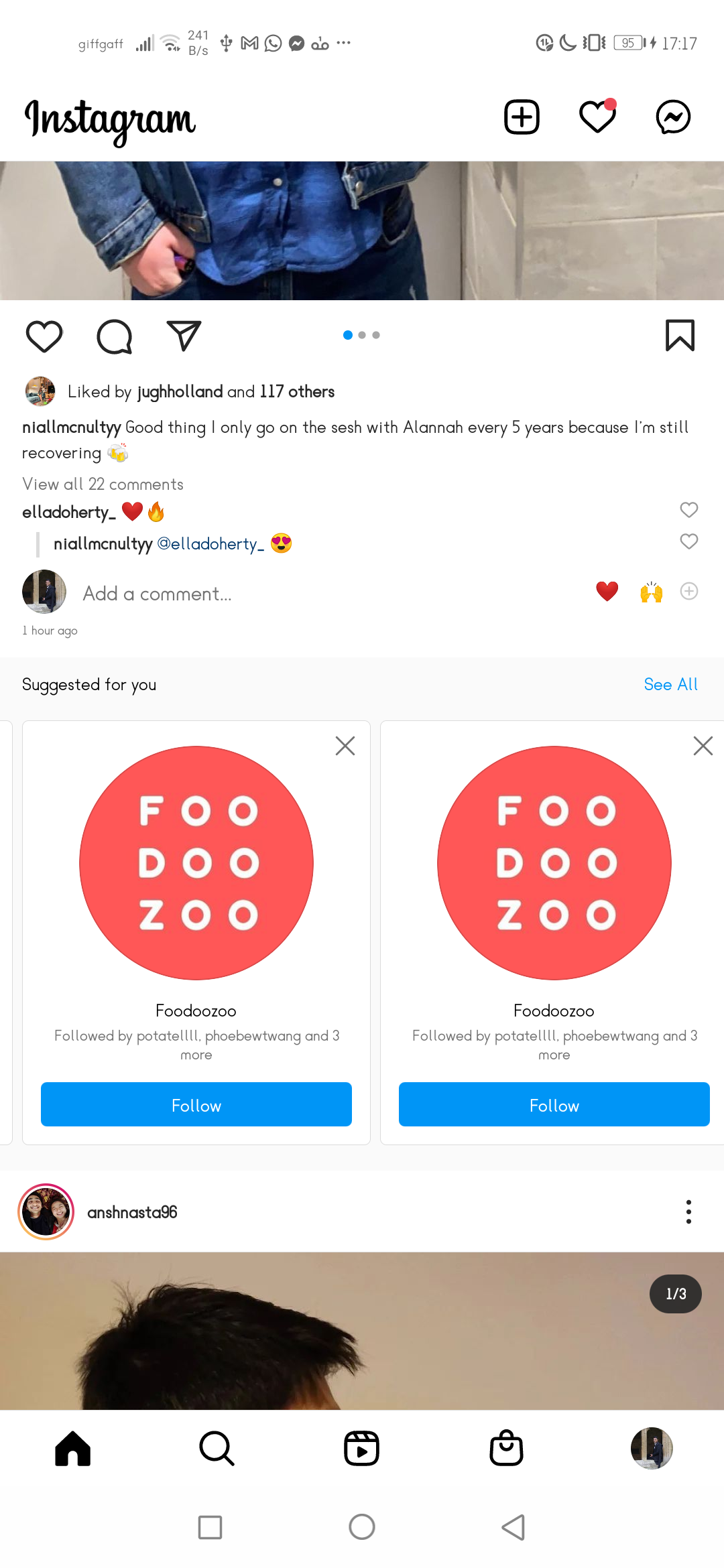}
    \includegraphics[height=4.8cm]{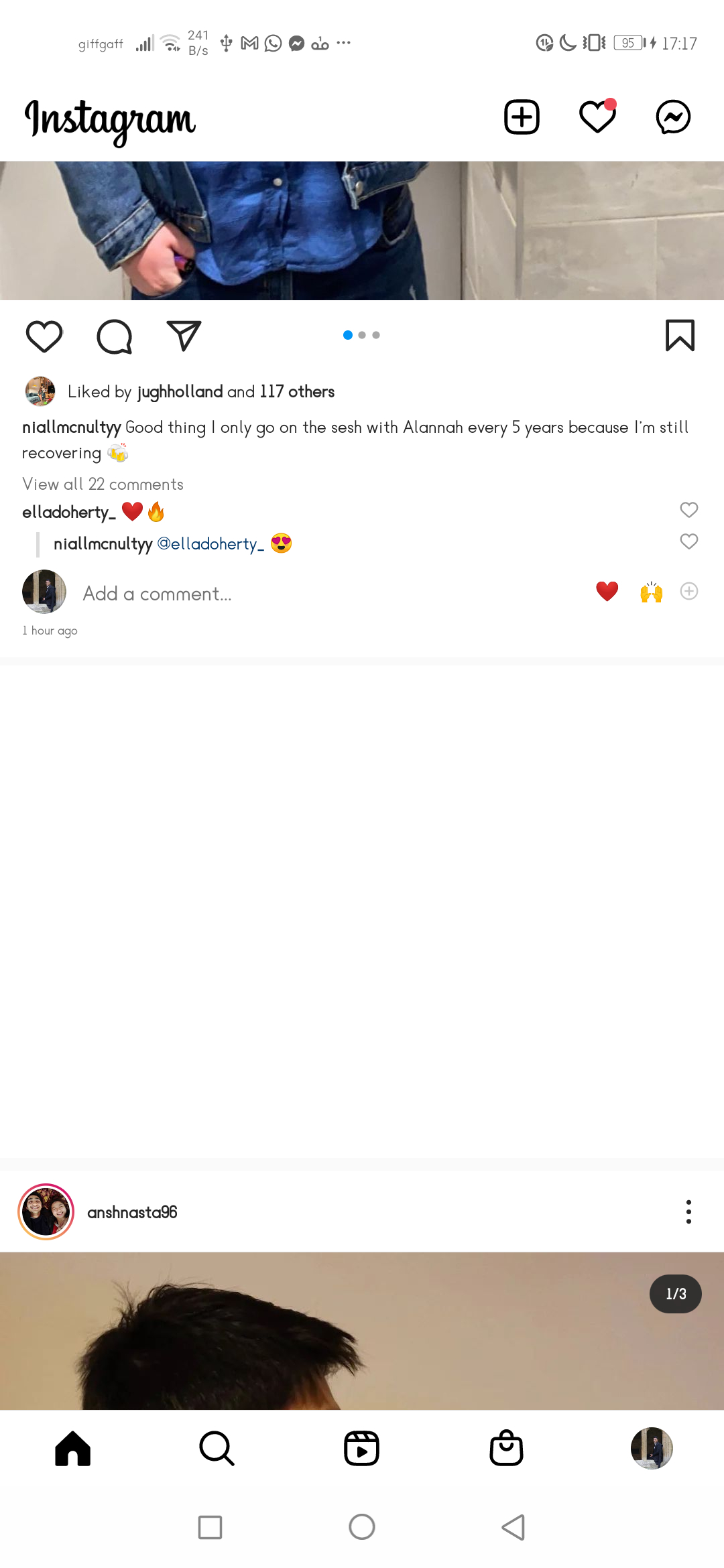}
    }
    \hfill
    \subfigure[Text censoring (YouTube \textit{left}, Reddit \textit{right})]{
    \includegraphics[height=4.8cm]{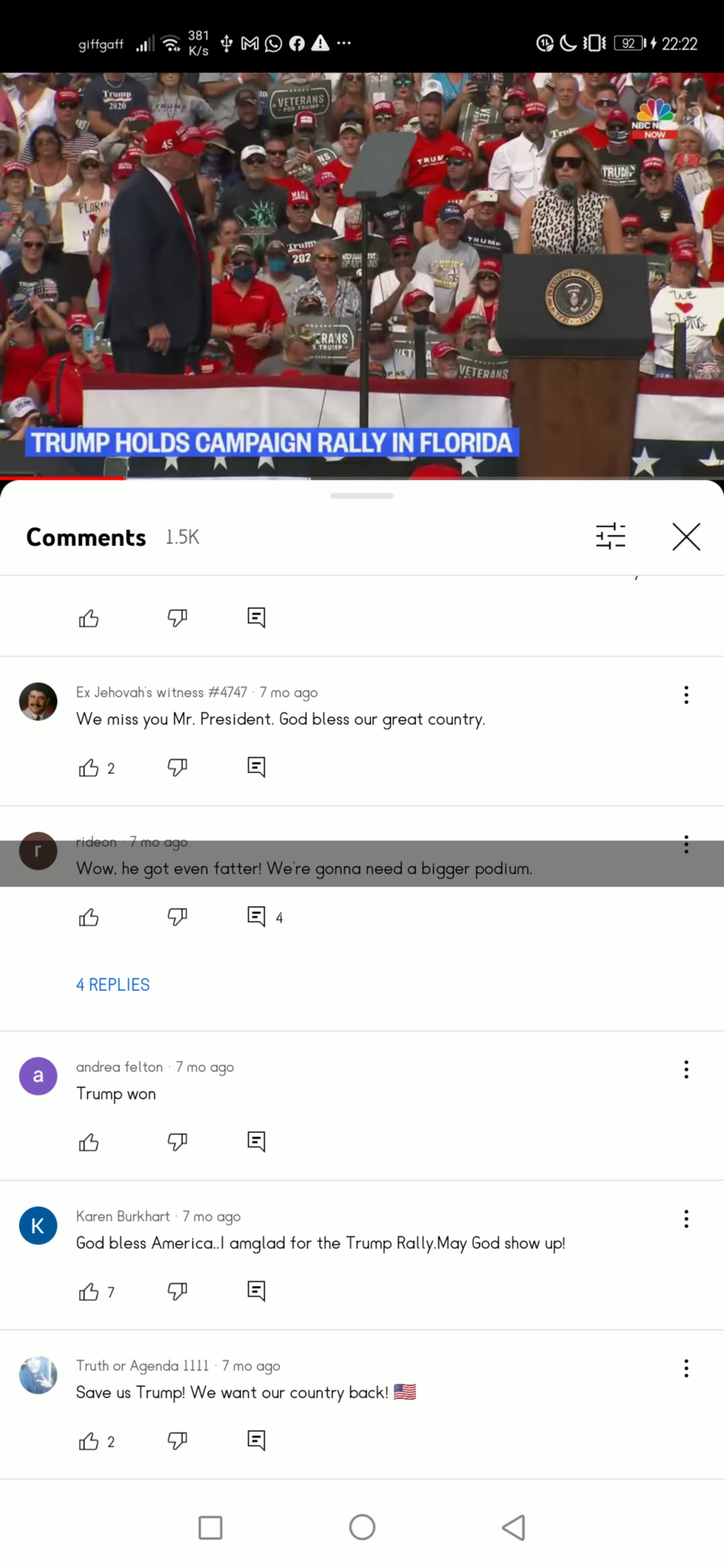}
    \includegraphics[height=4.8cm]{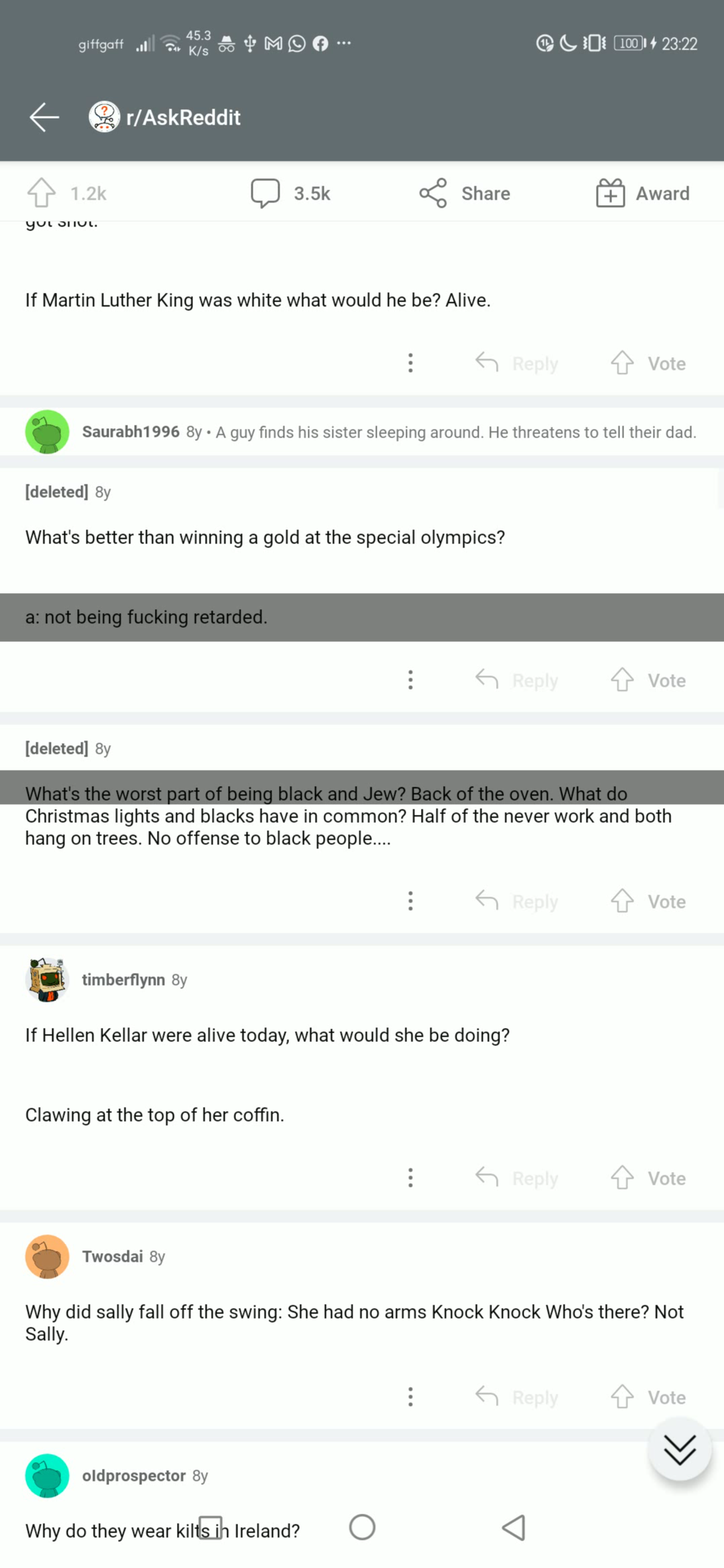}
    }
    \hfill
    \subfigure[Content moderation (Google Images, TikTok, YouTube, YouKu)]{
    \includegraphics[height=4.8cm]{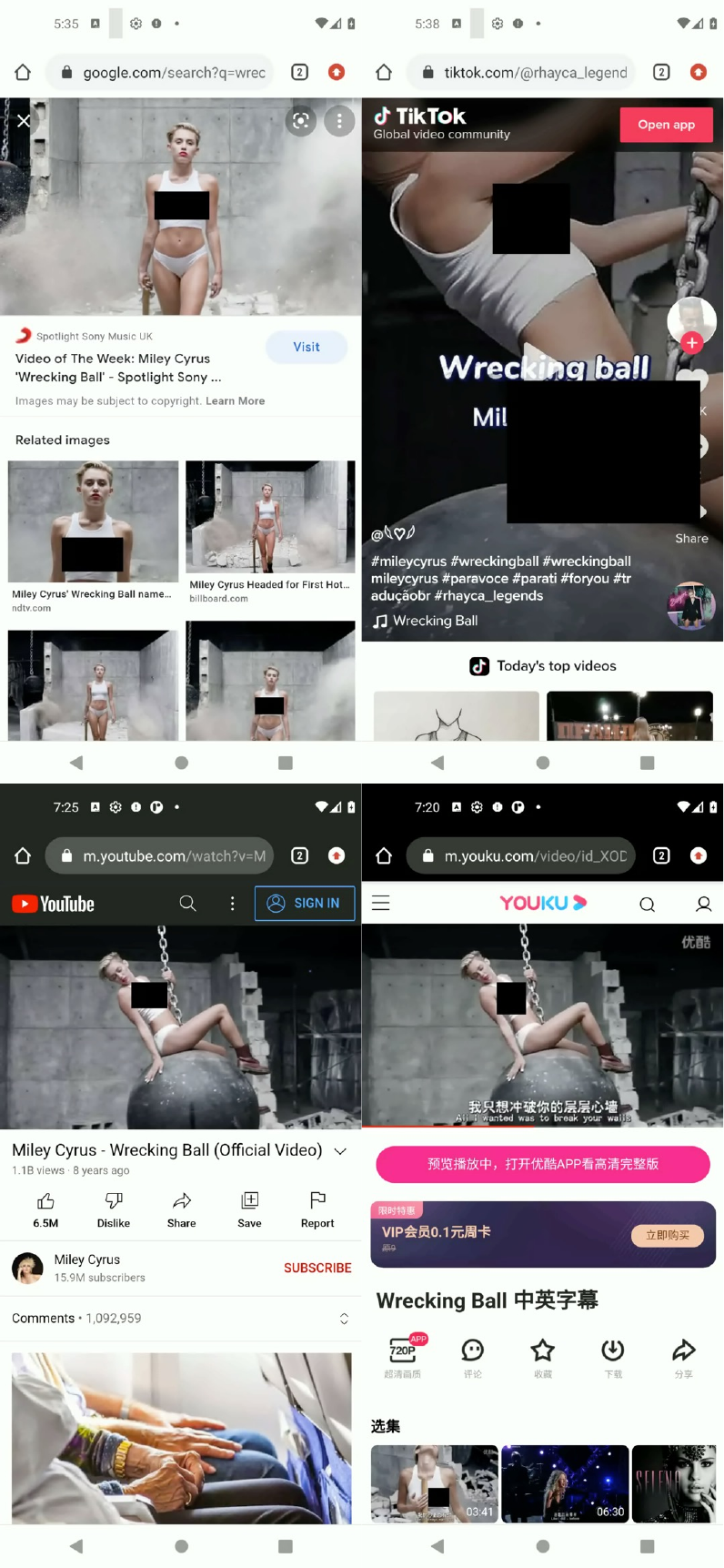}
    }
    \label{fig:gt_eval}
\end{figure*}

\subsection{Opportunities for Low-code Development in Interface Modification}


\textbf{Low-code development platforms} have been defined, according to practitioners, to be (i) low-code (negligible programming skill required to reach endgoal, potentially drag-and-drop), (ii) visual programming (a visual approach to development, mostly reliant on a GUI, and "what-you-see-is-what-you-get"), and (iii) automated (unattended operations exist to minimize human involvement) \cite{Luo2021CharacteristicsAC}.
Low-code development platforms exist for varying stages of software creation, from frontend (e.g. App maker, Bubble.io, Webflow), to workflow (Airtable, Amazon Honeycode, Google Tables, UiPath, Zapier), to backend (e.g. Firevase, WordPress, flutterflow); none exist for software modification of existing applications across interfaces. 
According to a review of StackOverflow and Reddit posts analysed by Luo et al. \cite{Luo2021CharacteristicsAC}, low-code development platforms are cited by practitioners to be tools that enable faster development, lower the barrier to usage by non-technical people, improves IT governance compared to traditional programming, and even suits team development; one of the main limitations cited is that the complexity of the software created is constrained by the options offered by the platform.  

User studies have shown that users can self-identify malevolent harms and habits upon self-reflection and develop desires to intervene against them \citep{10.1145/3479600, 10.1145/3313831.3376672}. 
%
Not only do end-users have a desire or interest in self-reflection,
but there is indication that end-users have a willingness to act.
Statistics for content violation reporting from Meta show that in the Jan-Jun 2021 period,
$\sim$ 42,200 and $\sim$ 5,300 in-app content violations were reported on Facebook and Instagram respectively \citep{fbtrans}
(in this report, the numbers are specific to violations in local law, so the actual number with respect to community standard violatons would be much higher; the numbers also include reporting by governments/courts and non-government entities in addition to members of the public). 
Despite a willingness to act,
there are limited digital visualization or reflection tools that enable flexible intervention development by end-users.
There are visualization or reflection tools on browser and mobile that allow for reflection (e.g. device use time \citep{10.1145/2968219.2971591}), 
and there are separate and disconnected tools for intervention (Section 2.2),
but there are limited offerings of flexible intervention development by end-users,
where end-users can observe and analyze their problems while generating corresponding fixes,
which thus prematurely ends the loop for action upon regret/reflection.
There is a disconnect between the harms analysis ecosystem and interventions ecosystem.
A barrier to binding these two ecosystems is the existence of low-code development platforms for end-users. While such tooling may exist for specific use cases on specific interfaces 
(e.g. web/app/game development) 
for mostly creationary purposes, there are limited options available for modification purposes of existing software, the closest alternative being extension ecosystems \citep{gd, chromewebstore}.
Low-code development platforms are in essence "developer-less", removing developers out of the software development/modification pipeline by reducing the barrier to development/modification through the use of GUI-based features and negligible coding, to the extent that an end-user can self-develop without expert knowledge.

\textbf{Human-in-the-Loop (HITL)} learning is the procedure of integrating human knowledge and experience in the augmentation of machine learning models.
It is commonly used to generate new data from humans or annotate existing data by humans.
Wallace et al. \citep{wallace-etal-2019-trick} constructed a HITL system of
an interactive interface where a human talks with a machine 
to generate more Q\&A language and train/fine-tune Q\&A models.
Zhang et al. \citep{10.1145/3292500.3330773} proposed a HITL system
for humans to provide data for entity extraction,
including requiring humans to
formulate regular expressions and highlight text documents,
and annotate and label data.
For an extended literature review, 
we refer the reader to Wu et al. \citep{https://doi.org/10.48550/arxiv.2108.00941}.
Beyond lab settings, 
HITL has proven itself in wide deployment, 
where a wide distribution of users have indicated a willingness and ability to perform tasks on a HITL annotation tool, \textit{reCAPTCHA}, to access utility and services.
In 2010, Google reported over 100 million reCAPTCHA instances are displayed every day \citep{recaptcha}
to annotate different types of data, 
such as deciphering text for OCR of books or street signs, 
or labelling objects in images such as traffic lights or vehicles.

While HITL formulates the structure for human-AI collaborative model development, 
\textbf{model fine-tuning} and \textbf{few-shot learning} formulate the algorithmic methods of adapting models to changing inputs, environments, and contexts.
Both adaptation approaches require the model to update its parameters with respect to the new input distribution.
For model fine-tuning, the developer re-trains a pre-trained model on a new dataset. This is in contrast to training a model from a random initialization. 
Model fine-tuning techniques for pre-trained foundation models, that already contain many of the pre-requisite subnetworks required for feature reuse and warm-started training on a smaller target dataset, have indicated robustness on downstream tasks \citep{galanti2022on, abnar2022exploring, NEURIPS2020_0607f4c7}.
If there is an extremely large number of input distributions
and few samples per distribution (small datasets),
few-shot learning is an approach where
the developer has separately trained a meta-model that learns how to change model parameters with respect to only a few samples. 
Few-shot learning has demonstrated successful test-time adaptation in updating model parameters with respect to limited test-time samples in both image and text domains \citep{Raghu2020Rapid, Koch2015SiameseNN, finn2017modelagnostic, datta2021learnweight}.
Some overlapping techniques even exist between few-shot learning and fine-tuning, such as constructing subspaces and optimizing with respect to intrinsic dimensions \citep{aghajanyan-etal-2021-intrinsic, datta2022low, 9157772}.

The raw data for harms and required interface changes
reside in the history of interactions between the user and the interface.
In the Screenome project \citep{e1b52b8c74c04b46932c8b36e8880c0e, doi:10.1080/07370024.2019.1578652}, the investigators proposed the study and analysis of the moment-by-moment changes on a person’s screen, by capturing screenshots automatically and unobtrusively every $t=5$ seconds while a device is on.
This record of a user's digital experiences represented as a sequence of screens that they view and interact with over time is denoted as a user's \textbf{screenome}.
Though not mobilized widely amongst users for their self-reflection or personalized analysis, 
integrating screenomes into an interface modification framework can play the dual roles of 
visualizing raw (harms) data to users
while
manifesting as parseable input for visual overlay modification frameworks.

\section{Scope}
\label{sec:scope}

We state early on the scope of our problem, solution, and evaluation. 
While further extension is possible, these scopes present the required constraints needed to evaluate our contributions, and are not intended as limitations. 
As the field of interface modification is broad, we narrow our scope onto digital harms, and specify the limits of its evaluation in Section 3.1.
With respect to the challenges and opportunities shared in Section 2, and the bounds set in Section 3.1, we define in Section 3.2 the system requirements which we wish to instill into \grease{}.
With respect to the system requirements in Section 3.2, 
we discuss under this same context the methods that would optimally evaluate these system requirements in Section 5.

\subsection{Problem Scope}

While the proposal for a collaborative low-code development system for end-users can have many use cases, 
such as customizing themes or UI/UX modernization of legacy interfaces, 
we extend on the developing literature on digital harms mitigation, in-line with prior work on interface re-rendering \citep{greaseterminator, gd}. 
Further on digital harms, we propose a methodology for end-users to construct their own interventions. 
We reference existing intervention designs that have been previously-evaluated (e.g. user studies), hence we do not replicate these studies to evaluate our reproduced designs.
The methodology for constructing a wide range of existing intervention designs with overlay graphics rendering has been validated in the \textit{GreaseTerminator} work \citep{greaseterminator}.
The scope of types of intervention development is not under consideration when evaluating this HITL approach.
We focus on interface occlusion, as it has been indicated in literature to be a sufficiently-strong source of interventions. 
It has been shown in the \textit{GreaseTerminator} work that other interventions can be crafted (e.g. screen-locking based on app usage), thus other output modalities such as interface augmentation (e.g. interpolating text or content from email or message apps over inpainted content rather than just inpainting with single colour, or interaction flow detection with RICO dataset \citep{10.1145/3126594.3126651}) is feasible, but not the focus of this work.
Though we retain mask removal from \textit{GreaseTerminator} and extend model hooks with fine-tuning and few-shot learning, 
we provide evaluation results specifically on element removal and text censoring.
Existing literature already review that few-shot learning is feasible and scalable for multiple modalities, including text and images.
%
Masks and text models
are also the optimal candidates of demonstrable few-shot learning and model fine-tuning for this HITL setting. 
We do not evaluate the models or training regimes themselves; pre-trained and open-sourced models are available online, and text classification (and hate speech detection), out-of-distribution detection, model fine-tuning, and few-shot learning are active research areas.
As we do not deploy the tool widely amongst end-users, 
we do not provide analysis on the distribution of harms and interventions from constructing a harms database/repository.
We do not evaluate the user interface of the HITL annotation tool as the layout has been user-evaluated and is considered a standard baseline interface in HITL \citep{labelme}. 

\subsection{System Requirements}

The goal of this paper shall be to construct a system/architecture that is developer-less and solely user-driven in interface modifications. In this system, users shall individually or collaboratively in small networks develop changes to their target user interface. 
The goal is evaluated, with constraint to the scope of our study. 

We derive the technical requirement (Requirement 1) and systemic requirement (Requirement 2) from our background in Section 2.
Section 2.1 shows that there is a wide harms landscape and distribution afflicting users, but a disconnected interventions development process by developers.
Section 2.2 and 2.3 shows that we can potentially scale intervention development by 
making intervention development \textit{interface-agnostic} and \textit{developer-less}. 
We identify that an individual end-user should be permitted to autonomously personalize their digital interfaces against the wide distribution of harms that can occur, 
despite the technical challenges an intervention developer and end-user faces regarding interface modification (e.g. changing code, escalating privilege, non-generalizable across version and/or app changes).
In parallel to these challenges, we find opportunities for both the individual user and a network of users to benefit from HITL and model adaptation.
Requirement 1 is focused on the usability and feasibility of \grease{} for an individual user; Requirement 2 is focused of that for a network of users.
Through Requirements 1 and 2, 
our work intends 
to bind the harms landscape to the intervention development process
through \grease{}.

\begin{enumerate}
    \item \textbf{Requirement 1:} 
    \textit{A complete feedback loop between user input (train-time) and interface re-render (test-time).}
    \item \textbf{Requirement 2:} 
    \textit{Prospects for scalability across the distribution of interface modifications (with respect to both harms landscape and rendering landscape).}
\end{enumerate}
\newpage

\section{\grease{}}
\label{sec:grease}

\begin{figure*}
    \centering
    \caption{Walkthrough of using \grease{}-modified interfaces.}
    \Description{Interaction flow of using \grease{}-modified interfaces.}
    \subfigure[\textit{User authentication}: Secure gateway to the user's screenomes, personal devices, and intervention development suite. ]{
    \includegraphics[height=4.5cm]{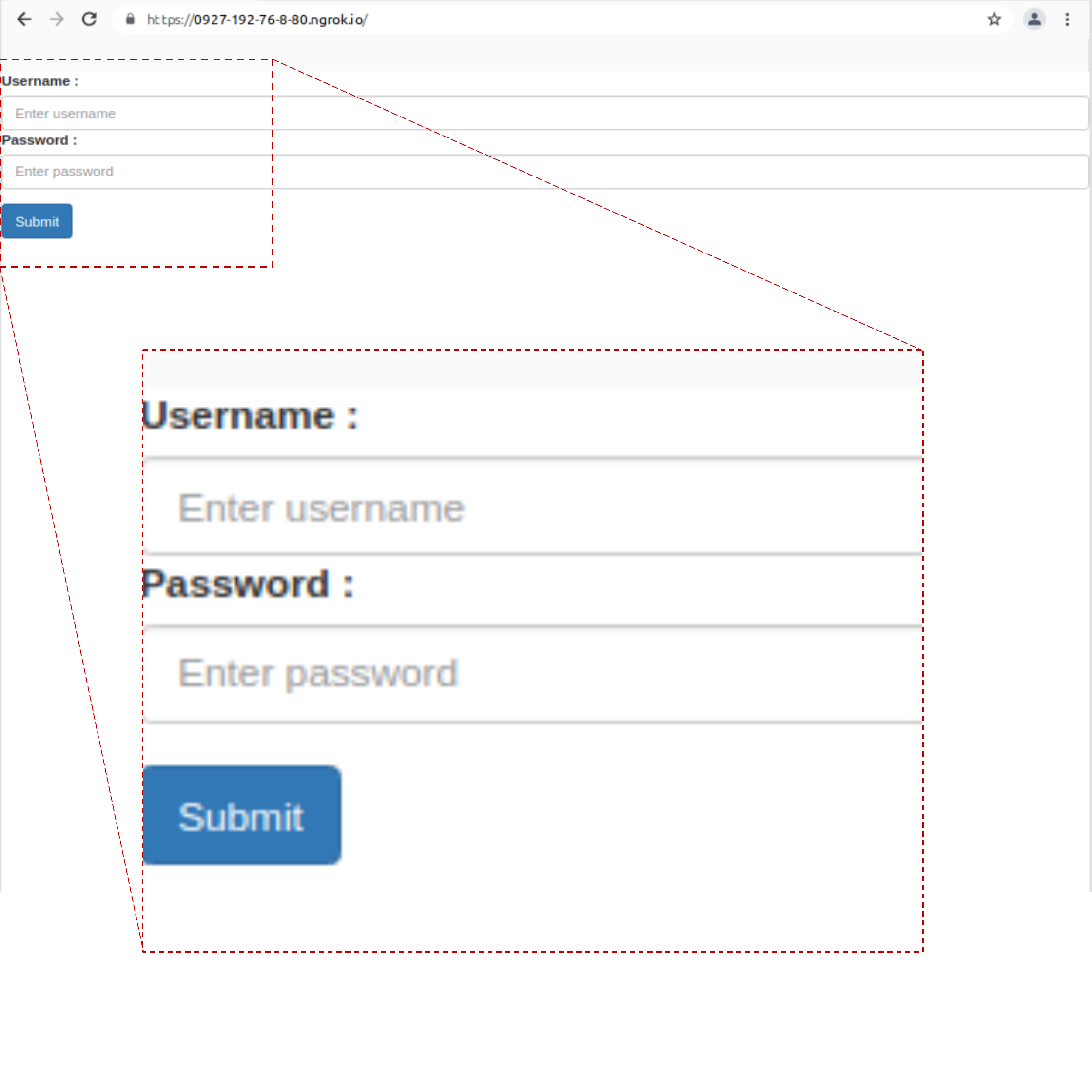}
    }\hfill
    \subfigure[\textit{Interface \& interventions selection}: Listings of all registered devices/emulators on server, as well as interventions contributed by the users or community using the tool in Figure \ref{fig:tagger}.]{
    \includegraphics[height=4.5cm]{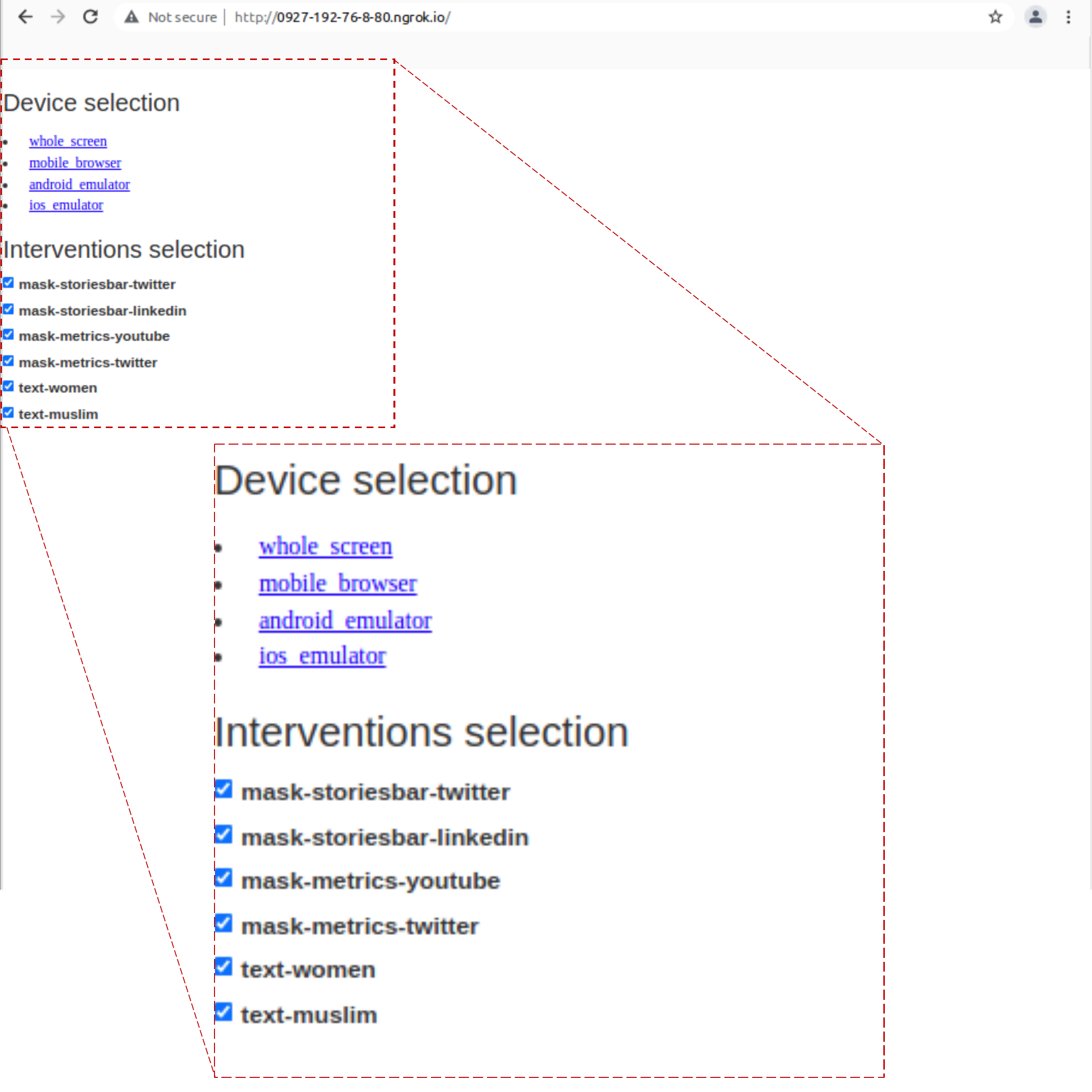}
    }\hfill
    \subfigure[\textit{Interface access}: Accessing a Linux desktop from another (Linux) desktop browser.]{
    \includegraphics[height=4.5cm]{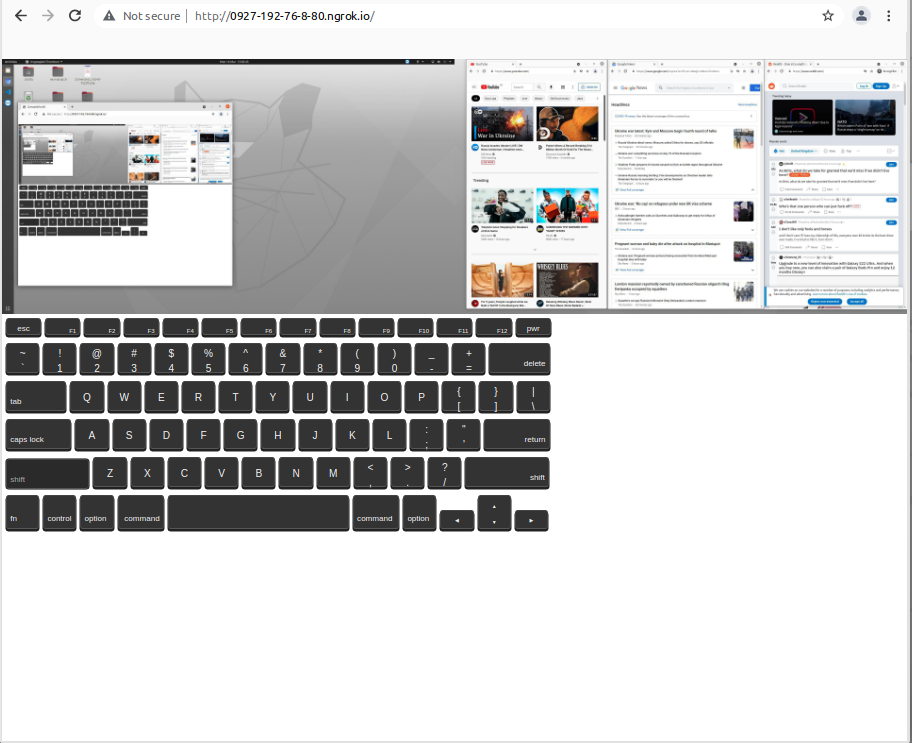}
    }
    \hspace{0.1cm}
    \subfigure[\textit{Interface access}: Accessing an Android emulator from another Android host device.]{
    \includegraphics[height=4.5cm]{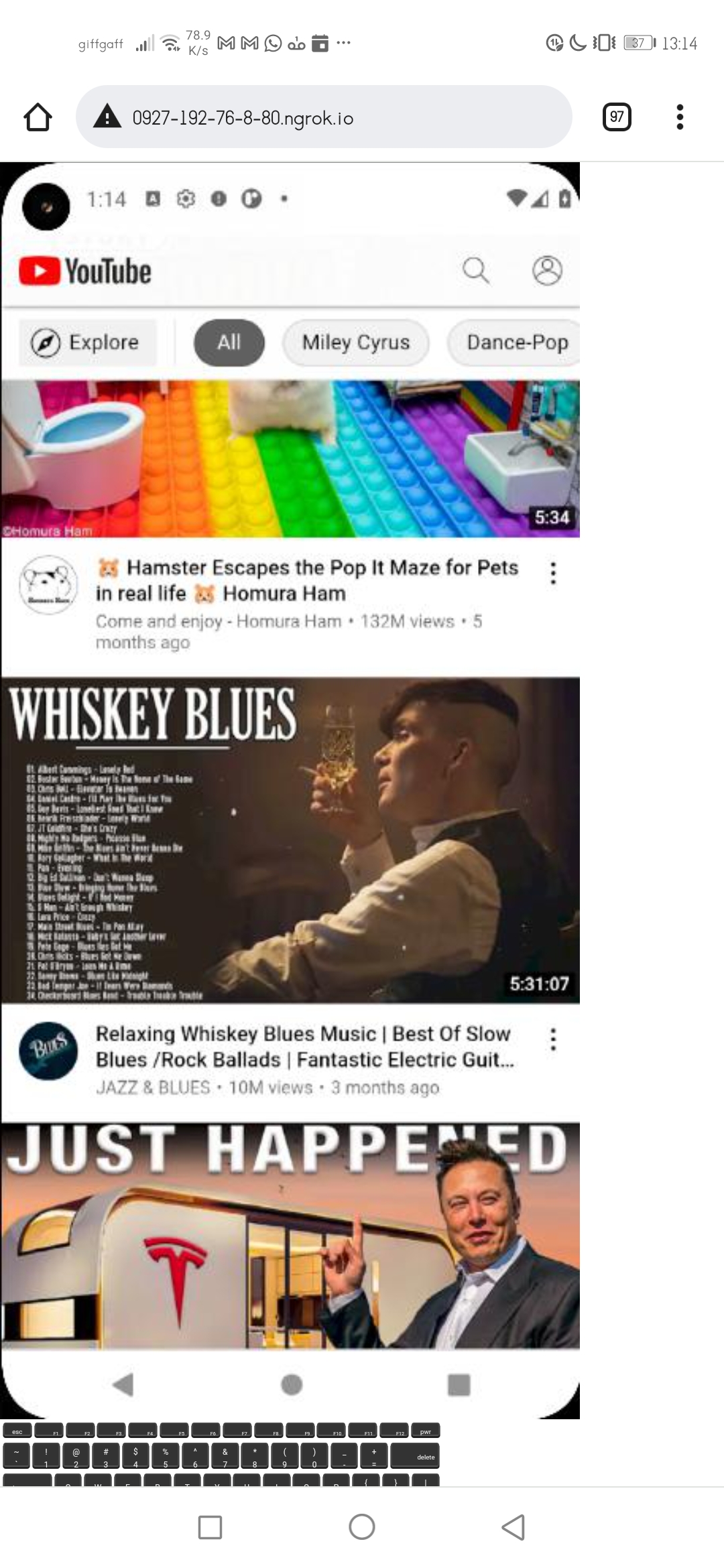}
    }
    \label{fig:walkthrough}
\end{figure*}



\subsection{System Architecture: Binding the Harms Ecosystem to the Interventions Ecosystem}

We define the \grease{} architecture, with which end-users (system administrators) interact with, as follows (Figure \ref{fig:archi_gw}(b)):
(i) the user logs into the \grease{} system to access amongst a set of personal emulators and interventions (the system admin has provisioned a set of emulated devices, hosted on a server through a set of virtual machines or docker containers for each emulator/interface, and handling streaming of the emulators, handling pre-requisites for the emulators, handling data migrations, etc); 
(ii) the user selects their desired interventions and continues browsing on their interfaces; 
(iii) after a time period, the user accesses their screenome and annotates interface elements, graphics, or text that they would like to generate interventions off of, which then re-populate the list of interventions available to members in a network. 

In our current implementation, the user accesses a web application (compatible with both desktop and mobile browsers). With their login credentials, the database loads the corresponding mapping of the user's virtual machines/containers that are shown in the interface selection page. 
The central server carries information on accessing a set of emulated devices (devices loaded locally on the central server in our setup). Each emulator is rendered in docker containers or virtual machines where input commands can be redirected.
The database also loads the corresponding mapping of available interventions (generated by the user, or by the network of users) in the interventions selection page. The database also loads the screenomes (images of all timestamped, browsed interfaces) in the screenome visualization page. 
Primary input commands for both desktop and mobile 
are encoded,
including keystroke entry (hardware keyboard, on-screen keyboard), mouse/touch input (scrolling, swiping, pinching, etc); input is locked to the coordinates of the displayed screen image on the web app (to avoid stray/accidental input commands), and the coordinates correspond to each virtual machine/container's display coordinates. 
Screen images are captured at a configurable framerate (we set it to 60FPS), and the images are stored under a directory mapped to the user. Generated masks and fine-tuned models are stored under an interventions directory and their intervention/file access is also locked by mapped users.
Interventions are applied sequentially upon a screen image to return a perturbed/new image, which then updates the screen image shown on the client web app. 



The improvements of \grease{} with respect to \textit{GreaseTerminator} are two-fold: 
(i) improvements to the framework enabling end-user development and harms mitigation (discussed in detail in Sections 4.2, 4.3, 5 and 6), and
(ii) improvements to the technical architecture (which we discuss in this section).
Our distinctive and non-trivial technical improvements to the \textit{GreaseTerminator} architecture fall under namely latency, device support, and interface-agnosticity.
%
\textit{GreaseTerminator} requires the end-user device to be the host device, and overlays graphics on top. A downside of this is the non-uniformity of network latency between users (e.g. depending on the internet speed in their location) resulting in a potential mismatch in rendered overlays and underlying interface.
With \grease{}, 
we send a post-processed/re-rendered image once to the end-user device's browser (stream buffering) and do not need to send any screen image from the host user device to a server, 
thus there is no risk of overlay-underlay mismatch and we even reduce network latency by half.
Images are relayed through an HTTPS connection, with a download/upload speed $\sim 250$Mbps, and each image sent by the server amounting to $\sim 1$Mb).
The theoretical latency per one-way transmission should be 
$\frac{1 \times 1024 \times 8 \textnormal{bits}}{250 \times 10^6 \textnormal{bits/s}}$ = 0.033ms.
With each user at most requiring server usage of one NVIDIA GeForce RTX 2080, with reference to existing online benchmarks \citep{aib} the latency for 1 image (CNN) and text (LSTM) model would be 5.1ms and 4.8ms respectively. 
While the total theoretical latency for \textit{GreaseTerminator} is ($2 \times 0.033 + 5$), that of \grease{} is ($0.033 + 5$) = 5.03ms. 
%
Another downside of \textit{GreaseTerminator} is that it requires client-side software for each target platform.
There would be pre-requisite OS requirements for the end-user device, where only versions of \textit{GreaseTerminator} developed for each OS can be offered support (currently only for Android).
\grease{} streams screen images directly to a login-verified browser, allowing users to access desktop/mobile on any browser-supported device. 
Despite variations in the streaming architecture between \grease{} and \textit{GreaseTerminator}, 
the interface modification framework (hooks and overlays) are retained, 
hence interventions (even those developed by end-users) from \grease{} are compatible in \textit{GreaseTerminator}.
In addition to improvements to the streaming architecture to fulfil interface-agnosticity,
adapting the visual overlay modification framework into a collaborative HITL implementation further improves the ease-of-use for all stakeholders in the ecosystem.
End-users do not need to root their devices, find intervention tools or even self-develop their own customized tools.
We eliminate the need for researchers to craft interventions (as users self-develop autonomously) or develop their own custom experience sampling tools (as end-users/researchers can analyze digital experiences from stored screenomes).
We also eliminate the need for intervention developers to 
learn a new technical framework or learn how to fine-tune models.
Running emulators on docker containers and virtual machines on a (single) host server
is feasible, and thus allows for
the browser stream to be accessible cross-device without restriction, e.g. access iOS emulator on Android device, or macOS virtual machine on Windows device.  
Certain limitations are imposed on the current implementation, such as a lack of access to the device camera, audio, and haptics; however, these are not permanent issues, and engineered implementations exist where a virtual/emulated device can route and access the host device's input/output sources \citep{vx}.


\begin{figure*}[ht]
    \centering
    \includegraphics[width=\linewidth]{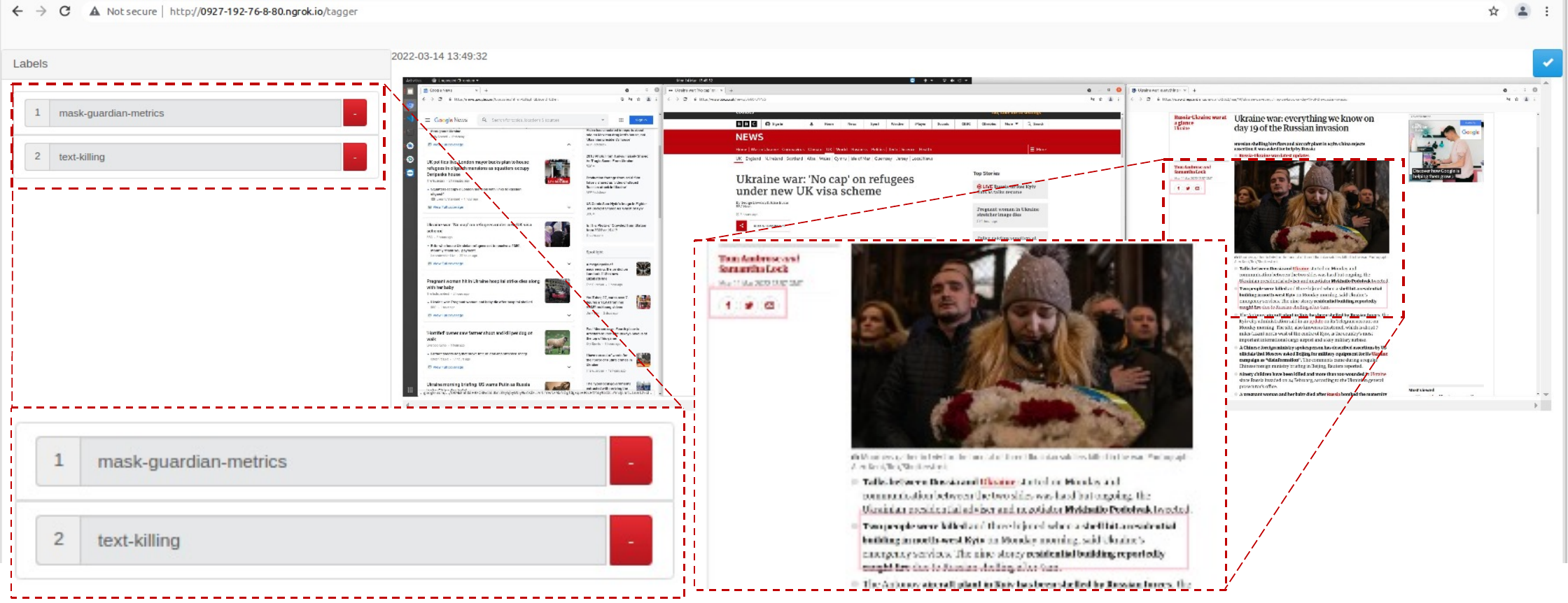}
    \caption{Screenome visualization page: The page offers the end-user the ability to traverse through the sequence of timestamped screen images which compose their screenome. They can use bounding boxes to highlight GUI elements, images or text. They can label these elements with specific encodings, such as \texttt{mask-} or \texttt{text-}.}
    \Description{Screenome visualization page: The page offers the end-user the ability to traverse through the sequence of timestamped screen images which compose their screenome. They can use bounding boxes to highlight GUI elements, images or text. They can label these elements with specific encodings, such as \texttt{mask-} or \texttt{text-}.}
    \label{fig:tagger}
\end{figure*}

\subsection{Interface Modification \& Re-rendering}


We make use of the three hooks from \textit{GreaseTerminator} (\textit{text}, \textit{mask}, and \textit{model} hooks), and link it with the screenome visualization tool. 
While in \textit{GreaseTerminator} the hooks ease the intervention development process for intervention developers with previous programming knowledge, 
we further generalize the intervention development process for intervention developers to the extent that even an end-user can craft their own interventions without developer support nor expert knowledge. \textit{GreaseTerminator} enables intervention generation (via hooks) and interface re-rendering (via overlays). The added \grease{} contribution of connecting these components with HITL learning and screenome visualization to replace developers is what exemplifies end-user autonomy and scalability in personalized interventions. 

The \textit{text hook} enables modifying the text that is displayed on the user's device.
It is implemented through character-level optical character recognition (OCR) that takes the screen image as an input and returns a set of characters and their corresponding coordinates. 
The EAST text detection~\citep{zhou2017east} model detects text in images and returns a set of regions with text, then uses Tesseract~\citep{tesseract} to extract characters within each region containing text.
The \textit{mask hook} matches the screen image against a target template of multiple images.
It is implemented with \textit{multi-scale multi-template} matching by resizing an image multiple times and sampling different subimages to compare against each instance of mask in a masks directory (where each mask is a cropped screenshot of an interface element). 
We retain the default majority-pixel inpainting method for mask hooks (inpainting with the most common colour value in a screen image or target masked region).
As many mobile interfaces are standardized or uniform from a design perspective compared to images from the natural world, this may work in many instances.
The mask hook could be connected to rendering functions such as highlighting the interface element with warning labels, or image inpainting (fill in the removed element pixels with newly generated pixels from the background), or adding content/information (from other apps) into the inpainted region.
Developers can also tweak how the mask hook is applied, for example using the multi-scale multi-template matching algorithm with contourized images (shapes, colour-independent) or coloured images depending on whether the mask contains (dynamic) sub-elements, or using few-shot deep learning models if similar interface elements are non-uniform. 
A \textit{model hook} loads any machine learning model to take any input and generate any output.
This allows for model embedding (i.e. model weights and architectures) to inform further overlay rendering.
We can connect models trained on specific tasks (e.g. person pose detection, emotion/sentiment analysis) to return output given the screen image (e.g. bounding box coordinates to filter), and this output can then be passed to a pre-defined rendering function (e.g. draw filtering box).

Recent developments in few-shot learning and model fine-tuning could potentially enable the scalability of interventions development through hooks and a low-code development platform.
\textit{GreaseTerminator} mask hooks tend to only require a single cropped image as input, and given the limited variability (or ease of re-cropping an updated interface design) of a specific GUI element on a given app, 1-shot of a GUI element is sufficient to detect it. 
Models used by \textit{GreaseTerminator} model hooks were expected to be either easily accessible from model zoos or model sharing platforms (e.g. PapersWithCode, Github, Kaggle, ModelZoo) or fine-tuned by intervention developers.
%
%
Incorporating model fine-tuning and few-shot learning mechanisms
into the interface modification framework can propagate new and personalized models/interventions for each user sub-group more efficiently without the need for a dedicated intervention developer to manually collect the required data themselves, if an open-ended input source is provided where an end-user can simply capture and submit new inputs over time. 
It reduces the feedback loop delay in deploying improved interfaces by removing all the development middlemen (including app/platform developers and external intervention developers).



\begin{figure*}
    \centering
    \caption{Removal of GUI elements (YouTube sharing metrics/buttons) across multiple target interfaces and operating systems. }
    \Description{Removal of sharing buttons}
    \subfigure[Element removal on emulated desktop (MacOS)]{
    \includegraphics[height=6cm]{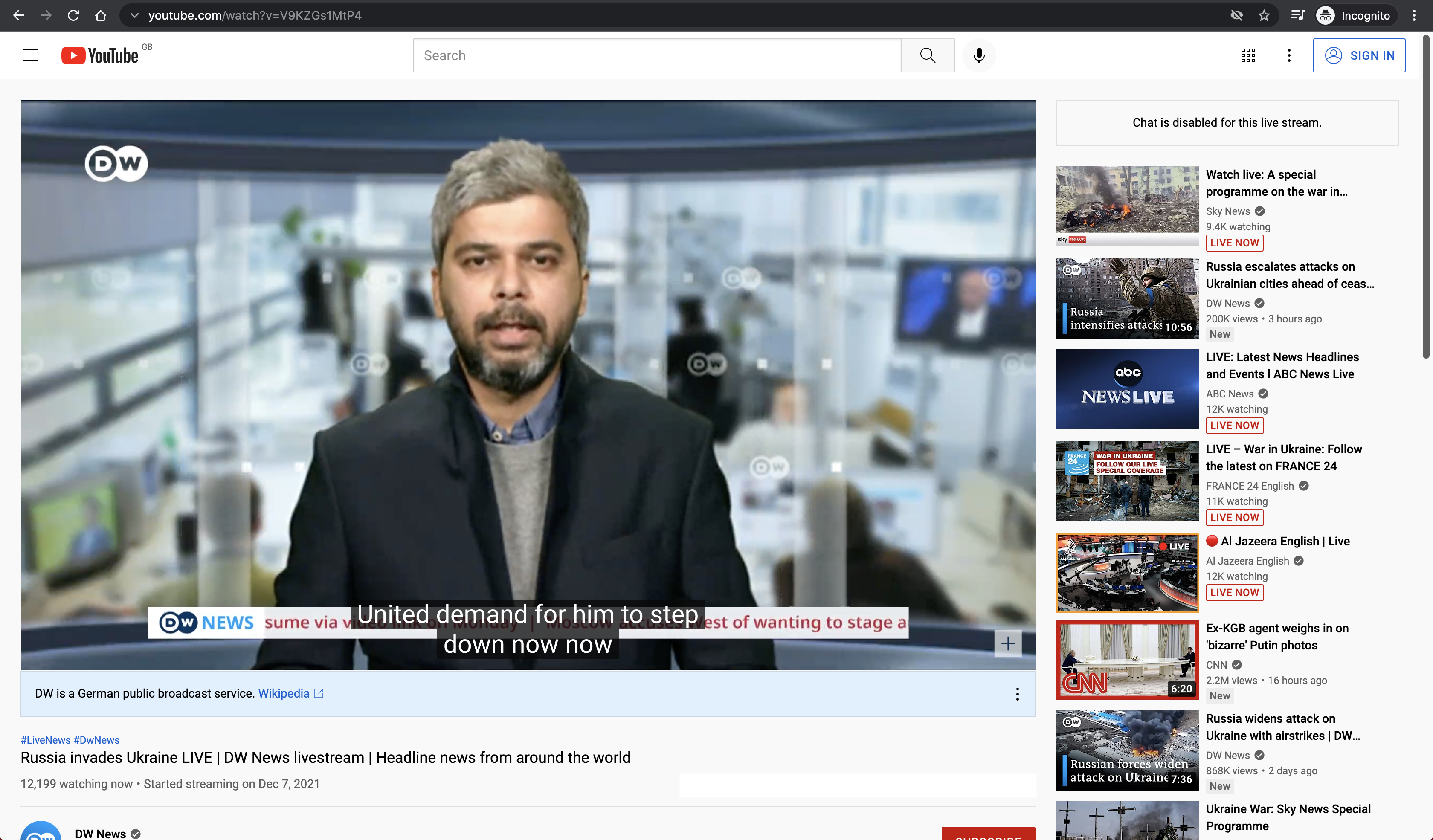}
    }\hfill
    \subfigure[Element removal on emulated Android]{
    \includegraphics[height=6cm]{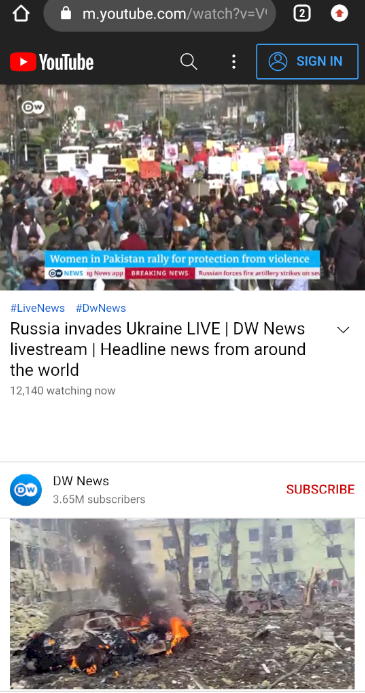}
    }\hfill
    \subfigure[Element removal on emulated iOS]{
    \includegraphics[height=6cm]{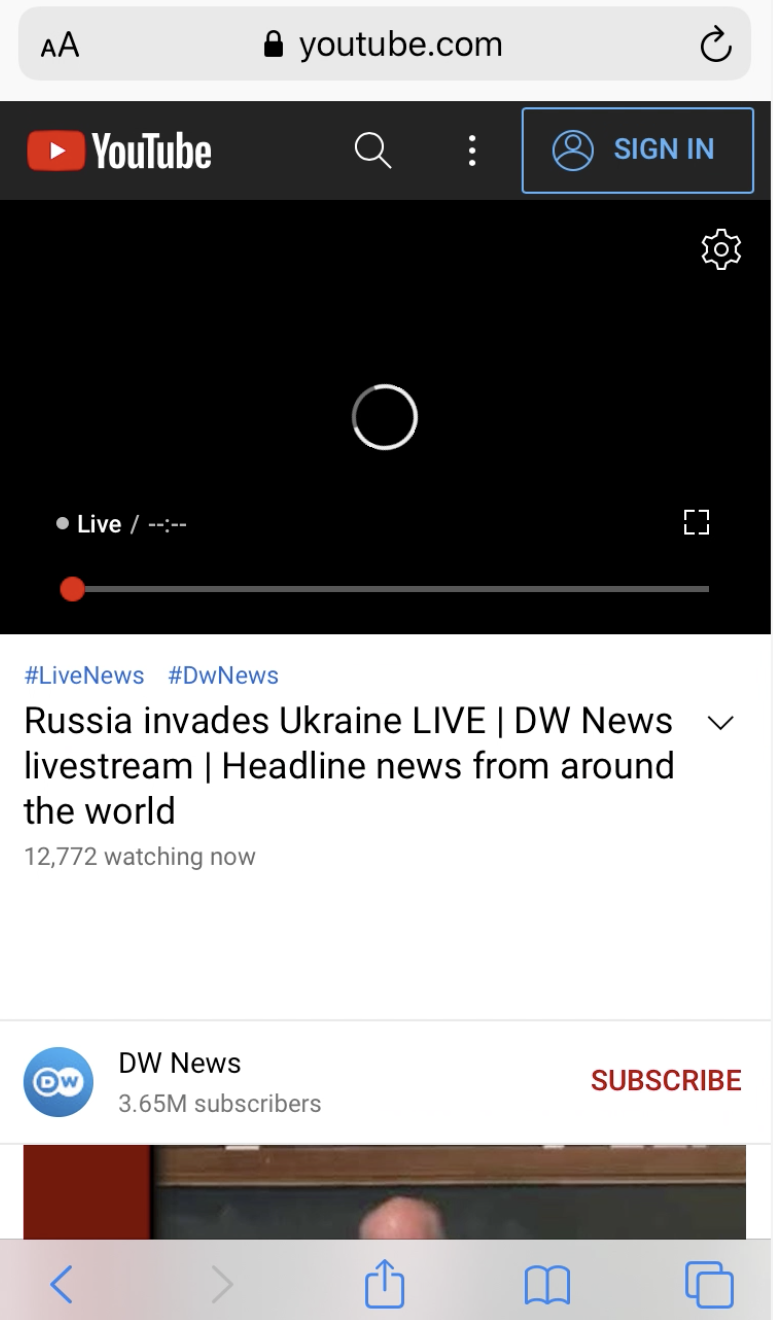}
    }
    \label{fig:sharebuttons_removal}
\end{figure*}

\subsection{Screenome Visualization \& Low-code Development}

An intersecting data source that enables both 
end-user self-reflection \citep{10.1145/3479600, 10.1145/3313831.3376672}
and interface re-rendering via overlay \citep{greaseterminator}
is the screen history or interface interaction history (denoted as \textit{screenome}).
Specifically, we can orchestrate a loop 
that receives input from users and generates outputs for users.
%
%
Through \grease{}, end-users can browse through their own screen history, and beyond self-analysis, they can constructively build interface modifications to tackle specific needs.

Extending on the interface rendering approach of overlays and hook-based intervention development, a generalizable design pattern for \textit{GreaseTerminator}-based interventions is observed, where current few-shot/fine-tuning techniques can reasonably approach many digital harms, given appropriate extensions to the end-user development suite. 
In the current development suite (Figure \ref{fig:tagger}), 
an end-user can inspect their screenomes across all \grease{}-enabled interfaces (ranging from iOS, Android to desktops), 
and make use of image segment highlighting techniques to annotate interface patterns to detect (typically UI elements or image/text) and subsequently intervene against these interface patterns.
Specifically, the interface images being stored and mapped to a user is shown in time-series sequence to the user. The user can go through the sequence of images to reflect on their browsing behavior. 
The current implementation focuses on one-shot detection of masks and fine-tuning of image and text classification models. When the user identifies a GUI element they do not wish to see across interfaces and apps, they highlight the region of the image, and annotate it as \texttt{mask-<name-of-intervention>}, and the mask hook will store a mask of intervention \texttt{<name-of-intervention>}, which will then populate a list of available interventions with this option, and the user can choose to activate it during a browsing session.
When a user identifies text (images) that they do not wish to see of similar variations, they can highlight the text (image) region, and annotate it as \texttt{text-<name-of-intervention>} (\texttt{image-<name-of-intervention>}). The text hook will extract the text via OCR, and fine-tune a pretrained text classification model specifically for this type of text \texttt{<name-of-intervention>}. For images, the highlighted region will be cropped as input to fine-tune a pretrained image classification model.
The corresponding text (image) occlusion intervention (or extensibly other augmentation interventions) will censor similar text (images) during the user's browsing sessions if activated.
Extending on model few-shot training and fine-tuning, we can scale the accuracy of the models, not just through improvements to these training methods, but also by improving the data collection dynamics. More specifically, based on the spectrum of personalized and overlapping intervention needs for a distribution of users, we can leverage model-human and human-human 
collaboration to scale the generation of mask and model interventions.
In the case of mask hooks, end-users who encounter certain harmful GUI elements (perhaps due to exposure to specific apps or features prior to other users) can tag and share the mask intervention with other users collaboratively.

To collaboratively fine-tune models, 
users tag text based on a general intervention/category label, that is used to group text together to form a mini-dataset to fine-tune the model. An example of this would be a network of users highlighting racist text they come across in their screenomes that made them uncomfortable during their browsing sessions, and tagging them as \texttt{text-racist}, which aggregates more sentences to fine-tune a text classification model responsible for detecting/classifying text as racist or not, and subsequently occluding the text for the network of users during their live browsing sessions.
%
The current premise is that users in a network know a ground-truth label of the category of the specific text they wish to detect and occlude, and the crowd-sourced text of each of $N$ categories will yield corresponding $N$ fine-tuned models.
Collaborative labelling scales the rate in which text of a specific category can be acquired, reducing the burden on a single user while also diversifying the fine-tune training set, while also proliferating the fine-tuned models across a network of users and not wasting effort re-training already fine-tuned models of other users (i.e. increasing scalability of crafting and usage of interventions). 
A concern with this assumption, is that the labelling of such inputs in the real-world may not be standardized, and similar inputs may be grouped separately or dissimilar inputs may be grouped together, if we purely rely on network-based tagging. 
To mitigate this, possible algorithmic approaches to ensuring similar texts are grouped together for fine-tuning, which in themselves are also active research areas and should not be used to constrain the evaluation of our system, would be the use of in/out-of-distribution detection (e.g. computing the interference in loss convergence with respect to 2 inputs coming from different categories, or using a similarity metric, in order to regroup contributed inputs into appropriate categories), or the use of ensemble models (e.g. preparing $M$ different batches of training sets to train $M$ different ensemble models, so that the dissimilarity between certain sentences do not afflict a single model alone, and other models can validate a prediction).
Additional extensions to the HITL interface
could be 
the visualization of statistics of existing interventions or newly-crafted interventions, 
such as 
tracked accuracy of the models, 
the aleatoric/epistemic uncertainty of an intervention (informing whether the intervention is suffering from insufficient data or intrinsic ambiguity in the data) \citep{10.1145/3490099.3511117},
or ratings from other users.

\section{Evaluation}
\label{sec:evaluation}


We describe the methods to evaluate our system against Section 3.2's requirements and 
share findings of this evaluation in Section 5, 
and discuss the implications of these findings in Section 6.
%
We evaluate the architectural design rather than our extensible technical implementation.
With respect to Section 3.2 requirements, we evaluate the usability of
(\textit{Req 1}) the HITL component (usability for a \textit{single} user with respect to inputs/outputs; or "does our system help generate interventions?"), and
(\textit{Req 2}) the collaborative component (improvement to usability for a \textit{single} user when \textit{multiple} users are involved; or "does our system scale with user numbers?").

We evaluate (\textit{Req 1}) and (\textit{Req 2}) with \textbf{cognitive walkthroughs} and \textbf{scalability tests}.
We explain the methodology used to evaluate each requirement, 
then we provide the raw data and information retrieved,
then provide findings from the data.
Evaluating with walkthrough demonstrations ("show and tell rather than use and test")
and
performance benchmarking,
while considering heuristics using a checklist approach,
are all adopted evaluation strategies in-line with suggestions made by Ledo et al. \citep{10.1145/3173574.3173610}.
We refer the reader to Strobelt et al. \citep{9552430} for an example of literature that evaluates a collaborative HITL system based on Ledo et al. \citep{10.1145/3173574.3173610}'s principles.

\subsection{Cognitive Walkthrough}

\begin{table*}[t]
\centering
	\begin{minipage}{0.45\textwidth}
		\centering
    \resizebox{\textwidth}{!}{
        \begin{tabular}{lccccc}
            \toprule
            Mask
            & Min. masks
            & Android app	
            & iOS app	
            & Mobile browser	
            & Desktop browser \\
            \midrule
            \multicolumn{6}{l}{\textit{Stories bar}} \\
            --- Twitter & 1 & \ding{51} & \ding{51} & --- & --- \\
            --- Linkedin & 1 & \ding{51} & \ding{51} & --- & --- \\
            --- Instagram & 1 & \ding{51} & \ding{51} & --- & --- \\
            \midrule
            \multicolumn{6}{l}{\textit{Metrics/Sharing bar}} \\
            - Facebook & 2 & \ding{51} & \ding{51} & \ding{51} & \ding{51} \\
            - Instagram & 2 & \ding{51} & \ding{51} & \ding{51} & \ding{51} \\
            - Twitter & 2 & \ding{51} & \ding{51} & \ding{51} & \ding{51} \\
            - YouTube & 2 & \ding{51} & \ding{51} & \ding{51} & \ding{51} \\
            - TikTok & 2 & \ding{51} & \ding{51} & \ding{51} & \ding{51} \\
            \midrule
            \multicolumn{6}{l}{\textit{Recommended items}} \\
            - Twitter & 2 & \ding{51} & \ding{51} & \ding{51} & \ding{51} \\
            - Facebook & 2 & \ding{51} & \ding{51} & \ding{51} & \ding{51} \\
            \bottomrule
            \vspace{0.2cm}
        \end{tabular}
        }
        \captionof{table}{\ding{51} if element removal is successful, \ding{55} if element removal is unsuccessful, --- if the element not available on an interface. 
        }
        \Description{\ding{51} if successful, \ding{55} if unsuccessful, --- if not applicable. }
        \label{tab:mask_results}
	\end{minipage}
	\hspace{1cm}
	\begin{minipage}{0.45\textwidth}
        \includegraphics[width=\textwidth]{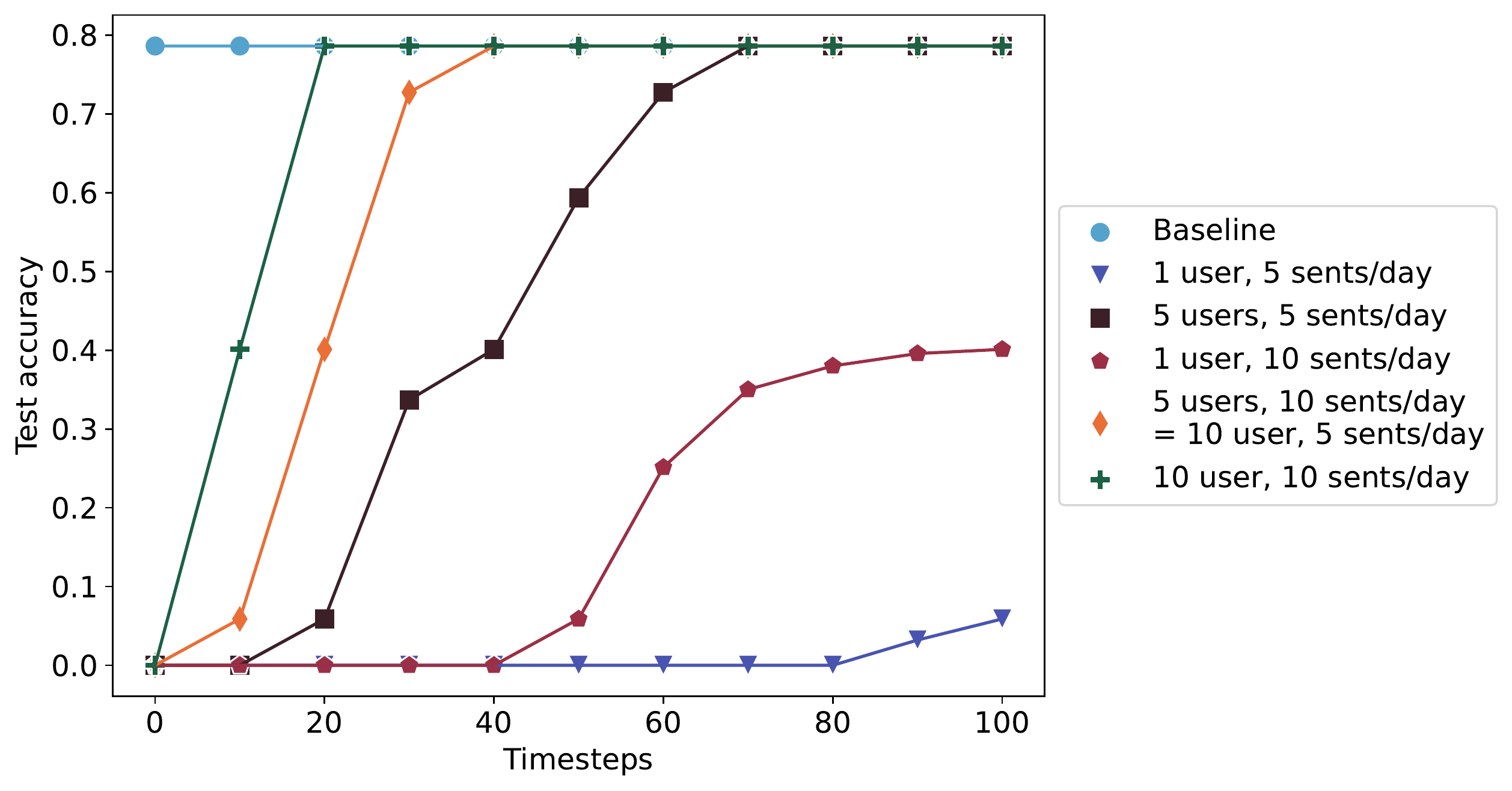}
        \captionof{figure}{Convergence of few-shot/fine-tuned models on sub-groups of hate speech}
        \Description{Convergence of few-shot/fine-tuned models on sub-groups of hate speech}
        \label{fig:fewshot_results} 
	\end{minipage}
\end{table*}


To evaluate the HITL component, 
rather than evaluating the output (interventions), we will evaluate the HITL process. 
Qualitatively, we perform a cognitive walkthrough of the user experience
%
to simulate the cognitive process and explicit actions taken by an end-user during usage of \grease{} to access interfaces and craft interventions.
The evaluation of the development process with respect to development cognition and development actions is based on work in cognitive walkthroughs~\citep{10.1145/223904.223962, 10.1145/223355.223735} and software engineering ethnographic studies~\citep{10.1145/3338906.3338976}.
%
%
%
While viewing their screenome,
as the time between when a user determines a set of pixels to be harmful and that of crafting a bounding box is negligible, rather than demonstrating the ease of intervention development by time spent, we can quantitatively supplement with the number of variations of interventions generated (specifically element removal) (Table \ref{tab:mask_results}). 

In our walkthrough, we as the authors/researchers presume the role of an end-user. 
We state the walkthrough \textbf{step} in \textbf{bold}, \textit{data pertaining to the task} in \textit{italics}, and descriptive evaluation in normal font. 
To evaluate the process of constructing an intervention using our proposed HITL system, based on the evaluation scope discussed, 
we evaluate \textit{the completion of a set of required tasks} based on criteria from
Parasuraman et al.'s \citep{844354} 4 types of automation applications, which aim to measure 
the role of automation in the harms self-reflection and intervention self-development process.
The four required tasks to be completed are:
\begin{enumerate}
    \item \textit{Information Acquisition}: 
    Could a user collect new data points to be used in intervention crafting?
    \item \textit{Information Analysis}:
    Could a user analyze interface data to inform them of potential harms and interventions?
    \item \textit{Decision \& Action Selection}: 
    Could a user act upon the analyzed information about the harms they are exposed to, and develop interventions?
    \item \textit{Action Implementation}: 
    Could a user deploy the intervention in future browsing sessions?
\end{enumerate}




\textbf{User logs in (Figure \ref{fig:walkthrough}a): }
\textit{The user enters their username and password. These credentials are stored in a database mapped to a specific (set of) virtual machine(s) that contain the interfaces the user registered for access. }
This is a standard step for any secured or personalized system, where a user is informed they are accessing data and information that is tailored for their own usage.

\textbf{User selects active interface and interventions (Figure \ref{fig:walkthrough}b): }
\textit{The user is shown a set of available interventions, be it contributed by themselves or other users in a network. They select their target interventions, and select an interface to access during this session. }
Based on their own configurations (e.g. \grease{} set up locally on their own computer, or specific virtual machines set up for the required interfaces), users can view the set of interfaces that they can access and use to facilitate their digital experiences. The interface is available 24/7, retains all their personal data and storage, is recording their screenome data for review, and accessible via a web browser from any other device/platform. They are less constrained by the hardware limitations of their personal device, and just need to ensure the server of the interfaces has sufficient compute resources to host the interface and run the interventions. 
The populated interventions are also important to the user, as it is a marketplace and ecosystem of personalized and shareable interventions. Users can populate interventions that they themselves can generate through the screenome visualization tool, or access interventions collaboratively trained and contributed by multiple members in their network. The interventions are also modular enough that users are not restricted to a specific combination of interventions, and are applied sequentially onto the interface without mismatch in latency between the overlay and underlying interface. As the capabilities of generating interventions (e.g. more hooks) and rendering interfaces (e.g. interface augmentation) become extended, so do their ability to personalize their digital experience, and generate a distribution of digital experiences to match a similarly wide distribution of users.
The autonomy to deploy interventions, with enhanced optionality through community-contributed interventions, before usage of an interface satisfies Task 4.

\textbf{The user accesses the interface and browses (Figure \ref{fig:walkthrough}c): }
\textit{The user begins usage of the interface through the browser from their desired host device, be it mobile or desktop. They enter input to the system, which is streamed to the virtual machine(s), and interventions render overlay graphics to make any required interface modifications. }
After the user has chosen their desired interventions,
the user will enjoy an improved digital experience through the lack of exposure to certain digital elements, such as undesired text or GUI elements. 
The altered viewing experience satisfies both Task 1 and 4; not only is raw screen data being collected, but the screen is being altered by deployed interventions in the wild.
The user cannot be harmed by what they previously chose not to see, and what they do see but no longer wish to see in the future, they can annotate to remove in future viewings in the screenome visualization tool. It is a cyclical loop where users can redesign and self-improve their browsing experiences through the use of unilateral or user-driven tools. 

\textbf{The user browses their screenome to generate interventions (Figure \ref{fig:tagger}): }
\textit{After a browsing period, the user may opt to browse and view their personal screenome. They enter the screenome visualization page to view recorded intervals of their browsing activity across all interfaces, and they can choose to annotate certain regions (image or text) to generate interventions to re-populate the interventions available. }
The user is given autonomy in selecting and determining what aspects of the interface, be it the static app interface of dynamic content provisioned, that they no longer wish to see in the future. 
Enabling the user to view their screenome across all used digital interfaces (extending to mobile and desktop) to self-reflect and analyze browsing or content patterns satisfies Task 2. 
Though the screenome provides the user raw historical data, it may require additional processing (e.g. automated analysis, charts) to avoid information overload.
Rather than waiting for a feedback loop for the app/platform developers or altruistic intervention developers to craft broad-spectrum interventions that may or may not fit their personal needs, the end-user can enjoy a personalized loop of crafting and deploying interventions, almost instantly for certain interventions such as element masks. The user can enter metadata pertaining to each highlighted harm, and not only contribute to their own experience improvement, but also contribute to the improvement of others who may not have encountered or annotated the harm yet.
By developing interventions based on their analysis, not only for themselves but potentially for other users,
they could successfully achieve Task 3. 
Though previously-stated as out of scope, to further support Task 3, other potential intervention modalities such as augmentation could also be contributed by a community of
professional intervention developers/researchers (who redirect efforts from individual interventions towards enhancing low-code development tools).

The four tasks, used to determine whether a complete feedback loop between input collection/processing and interface rendering through HITL by a single user, 
could all be successfully completed,
thus \grease{} satisfies Requirement 1.

\subsection{Scalability Testing}




To evaluate the collaborative component,
we measure the improvement to the user experience of a \textit{single} user through the efforts of multiple users.
We do not recruit a large number of real users, as it would bind our performance evaluation to the number of users available (which is additionally subject to constraints such as the evaluation period, intervention quality control, diversity of recruited users). 
Instead, we evaluate through scalability testing  \citep{stest}, a type of load testing that measures a system's ability to scale with respect to the number of users.

We simulate the usage of the system
to evaluate the scalable generation of one-shot graphics (mask) detection, and scalable fine-tuning/few-shot training of (text) models, 
in order to evaluate the strengths and weaknesses of the system's scalability.
We do not replicate the scalability analysis on real users: the fine-tuning mechanism is still the same, and the main variable (in common) is the sentences highlighted (and their assigned labels and metadata, as well as the quality of the annotations), though error is expectedly higher in the real-world as the data may be sampled differently and of lower annotation quality. 
The primary utility of collaboration to an individual end-user is the scaled reduction of effort in intervention development. 
We evaluate this in terms of variety of individualized interventions (variations of masks),
and the time saved in constructing a single robust intervention (time needed to construct an accurate model intervention).

\textbf{Breadth of interface-agnostic masks (Table \ref{tab:mask_results}): }
We investigate the ease 
to annotate graphically-consistent GUI elements for few-shot detection. 
We sample elements to occlude that can exist across a variety of interfaces.
We evaluate the occlusion of the \textit{stories bar} (pre-dominantly only found on mobile devices, not desktop/browsers); 
some intervention tools exist on Android \citep{swipe, InstaPrefs, gd, greaseterminator} and iOS \citep{friendly}, though the tools are app- (and version-) specific.
We evaluate the occlusion of \textit{like/share metrics}; there are mainly desktop browser intervention tools \citep{fbd, twd, igd, hidelikes}, and one Android intervention tool \citep{greaseterminator}.
We evaluate the occlusion of \textit{recommendations}; 
there are intervention tools that remove varying extents of the interface on browsers (such as the entire newsfeed) \citep{erad, unhook}.
Existing implementations and interest in such interventions indicate some users have overlapping interests in tackling the removal or occlusion of such GUI elements, though the implementations may not exist across all interface platforms, and may not be robust to version changes. 
For each intervention, we evaluate on a range of target (emulated) interfaces.
We aim for the real-time occlusion of the specific GUI element, and evaluate on native apps (for Android and iOS) and browsers (Android mobile browser, and Linux desktop browser). 

For each of the GUI element cases, we make use of the screenome visualization tool to annotate and tag the minimum number of masks of the specific elements we wish to block across a set of apps. There tend to be small variations in the design of the element between browsers and mobile, hence we tend to require at least 1 mask from each device type; Android and iOS apps tend to have similar enough GUI elements that a single mask can be reused between them. We tabulate in Table \ref{tab:mask_results} the successful generation and real-time occlusion of all evaluated and applicable GUI elements. 
We append screenshots of the removal of recommended items from the Twitter and Instagram apps on Android (Figure \ref{fig:gt_eval}(a,b)).
We append screenshots of the demetrification (occlusion of like/share buttons and metrics) of YouTube across 
desktop browsers (MacOS) and mobile browsers (Android, iOS)
(Figure \ref{fig:sharebuttons_removal}).

\textbf{Convergence of few-shot/fine-tune trained text models (Figure \ref{fig:fewshot_results}): }
We investigate the accuracy gains from fine-tuning pretrained text models as a function of user numbers and annotated sentence contributions. 
Specifically, we evaluate the text censoring of hate speech, where the primary form of mitigation is still community standard guidelines and platform moderation, with a few end-user tooling available \citep{bodyguard, greaseterminator}.
The premise of this empirical evaluation is that we have a group of simualated users $N$ who each contribute $N$ inputs (sentences) of a specific target class (hate speech, specifically against women) per timestep. 
With respect to a baseline, which is a pretrained model fine-tuned with all available sentences against women from a hate speech dataset, we wish to observe how the test accuracy of a model fine-tuned with $M \times N$ sentences varies over time. 
Our source of hate speech for evaluation is the \textit{Dynamically Generated Hate Speech Dataset} \citep{vidgen-etal-2021-learning}, which contains sentences of \texttt{non-hate} and \texttt{hate} labels, and also classifies hate-labelled data by the target victim of the text (e.g. \texttt{women}, \texttt{muslim}, \texttt{jewish}, \texttt{black}, \texttt{disabled}). As we expect the $M$ users to be labelling a specific niche of hate speech to censor, we specify the subset of hate speech of \texttt{women} (train set count: 1,652; test set count: 187).
We fine-tune a publicly-available, pre-trained \textit{RoBERTa} model \citep{roberta, DBLP:journals/corr/abs-1907-11692}, which was trained on a large corpus of English data (\textit{Wikipedia} \citep{wikidump}, \textit{BookCorpus} \citep{Zhu_2015_ICCV}).
For each constant number of users $M$ and constant sentence sampling rate $N$, at each timestep $t$, $M \times N \times t$ sentences are acquired of class \texttt{hate} against target \texttt{women}; there are a total of 1,652 train set sentences under these constraints (i.e. the max number of sentences that can be acquired before it hits the baseline accuracy), and to balance the class distribution, we retain all 15,184 train set \texttt{non-hate} sentences. We evaluate the test accuracy of the fine-tuned model on all 187 test set women-targeted hate speech. 
We also vary $M$ and $N$ to observe sensitivity of these parameters to the convergence towards baseline test accuracy.

The rate of convergence of a finetuned model is quicker when the number of users and contributed sentences per timestep both increase, approximately when we reach at least 1,000 sentences for the \texttt{women} hate speech category. 
The difference in convergence rates indicate that a collaborative approach to training can scale interventions development, as opposed to training text classification models from scratch and each user annotating text alone. 

The empirical results for this section are stated in Table \ref{tab:mask_results} and Figure \ref{fig:fewshot_results}.
The data and evaluations from the scalability tests indicate that the ease of mask generation and model fine-tuning, further catalyzed by performance improvements from more users, 
enable the scalable generation of interventions and their associated harms, thus \grease{} satisfies Requirement 2.

\section{Discussion}
\label{sec:discussion}

\subsection{What this means for understanding harms, interventions, and interfaces}

Software is composed of interface/graphics and functionality. 
Traditionally, the modification of software means changing source code, but we have shown some of the first steps towards illusory modifications, where we change the interface/graphics to change a user's perception of the software, which then changes how they use the software (i.e. perceived functionality change). 
Realistic functionality generation could be a next step,
through further study into generative interfaces \citep{74ffee915d6e4089bbe9f3ef3196ec8f, 10.1145/3313831.3376589}, generalization of server-less app functionality (e.g. learning from RICO dataset), program induction/synthesis from GUI \citep{10.1145/3220134.3220135}, and context-aware screen state prediction \citep{actionbert} and human behavioural prediction \citep{10.1145/3290607.3313089}. 

In the software development life cycle, maintenance is usually performed by the platform/software developers. Extension ecosystems extended upon this, by enabling intervention/patch developers to propose and share software patches, which may or may not be integrated into the next version update of the software. The proposed paradigm offers the notion that end-users themselves can takeover the maintenance and repair of their software, resulting in faster and needs-accurate version updates.

When users inspect and modify elements of the software and content, the resulting aggregated database of interface changes (harms and their mapped interventions) can benefit users themselves, other users in the software's ecosystem, software developers, and regulatory bodies.
The faster and public feedback cycle of what interface elements and content that varying portions of the user base like or dislike is informative for software regulation and policy-making, optimal software/interface design, optimal content moderation policies, sustainable digital habits, logging of harms and perpetrators, logging of unsustainable design practices, and so on. 

The collaborative nature of the proposed system reinforces the quality of interventions. The more users that share a specific need, the more data they contribute towards those interventions (e.g. tagging more elements or text). For niche users along the long-tail distribution, the interventions they craft are also of great importance, as they ensure the availability of niche interventions for themselves and other niche users that might have been previously ignored by intervention/software developers.

Enhancements and extensions to the current system are possible too.
Usage of app state exploration tools \citep{10.1145/2544173.2509552} and alternative GUI segmentation and parsing tools \citep{10.1145/3368089.3417940, 10.1145/3411764.3445049, 10.1145/3472749.3474763} can improve technical aspects of the implementation.
Usage of end-user local desktops as server, or loading personal devices instead of emulators or only using browsers to access apps (most popular apps also exist as websites), can allow for secure and private access to devices without trusting a third-party server to manage an end-users personal devices and data.
Rating of user mechanisms, detection of out-of-distribution input samples, or ensembling techniques are algorithmic techniques to handle the quality of collaborative interventions contributed.
Optimal intervention design may not be to pick one extreme (end-users should develop their own interventions) or the other (professional developers should develop interventions for end-users). We could balance the trade-offs to maximize the scalability of needs/harms and intervention design contributions from end-users, against fine-tuning and development skill from professional developers: end-users can define the requirements of the initial prototype of interventions through the low-code development platform and circulate for usage in the network, and 
professional developers can refine the intervention implementation 
based on criteria such as popularity of the intervention, amount of data contributed, and difficulty of re-development.




\subsection{The new standard of end-user autonomy}

We respond to the open call to action in Datta et al. \cite{greaseterminator} that asks for intervention development that increases autonomy for users while reducing digital paternalism. 
Datta et al. argued that a risk of digital paternalism manifests when interventions are developed by individuals other than those directly affected (e.g. researchers/developers building interventions without a sound understanding of the actual underlying challenges within an affected community).
As a solution to this potential risk, at best, the affected individuals should be able to develop interventions against harms \textit{by themselves}.

While \textit{GreaseTerminator} aimed to strike a balance between autonomy and paternalism \cite{Vandeveer2014} in regulating digital experiences, 
\grease{} aims to further increase end-user autonomy (and likewise reduce digital paternalism). 
We measure the extent of digital paternalism based on the extent in which an end-user delegates autonomy over their device screen to an external party. The complete trust of interface design and functionality to the original app developers is complete paternalism (negligible autonomy), given that the end-user is subject to the goals and decisions of the developer. The introduction of software modification frameworks for apps and browsers reduces paternalism slightly, in which the intervention developers may account for the end-users' best interests and goals, but it still requires trust from the end-user to the intervention developer. The proposed developer-less intervention development framework enables users to develop their own interventions (without expert knowledge) with respect to their own personalized goals, inducing maximum autonomy. However, this autonomy would still face constraints from the limitations of the development system, such as the generalizability of interface patterns, adverse network dynamics (e.g. other end-users over-contribute malicious samples to the few-shot learning models). With gradual improvements to implementation, this framework could be an optimal path to a non-paternalistic, autonomy-maximizing software modification framework. 

While this tool will mostly be used in small-scale studies, and further development will be required towards large-scale deployment, this first step towards end-user development for practical purposes highlights a potential tendency for the development landscape to change, shifting away from tooling, access, and efforts for/by professional developers, and more towards end-users. If other stakeholders oppose this movement, countermeasures may be imposed, such as technical restrictions (e.g. restricting usage of emulators, screen-image awareness, dynamic GUI elements) or legal restrictions (e.g. UI is copyrighted and not changeable). 
\newpage


\subsection{Re-writing digital realities}

With a system that allows for user-personalized, unilateral software modification such that the distribution of user interfaces match the distribution of user preferences, 
it inductively follows that end-users can personalize and re-write their "digital realities" (and by extension, their reality as a whole).

It has been argued in various forms that a coherent mesh exists between the physical reality and digital reality, the former being the physical world/reality/environment in which end-users exist and interact as physical human bodies, the latter being the digital world/reality/environment in which end-users exist and interact as digital personas. 
This coherent mesh is argued to exist, as phenomena occurring in the physical world can leak/impact the end-user manifestation in the digital world, and vice versa phenomena occurring in the digital world can leak/impact the end-user manifestation in the physical world.
An example is the leaking effects of online social networks onto the physical world: phenomena such as the viral propagation of content to large populations with or without filter, the large accessibility of other end-users around the world without a constrained contact list, the minimal or circumventable regulation of content generation resulting in real-world consequences such as the wide-scale re-influencing and recruitment of terrorists/extremists, wide-scale personalized influencing and biasing of niche populations for election rigging, or wide-scale social bubble formation of extreme interest groups (e.g. "incel", "sigma males", "furries") that lead to individuals exerting/manifesting these opinions in real-life against other individuals (e.g. exerting these personalities in the workplace or personal lives). 

One of the downsides of limited end-user autonomy and personalized interface rendering, is that the reality of the end-user has been shaped by platforms and other developers, and subsequently other end-users in the social network. 
If we aggregated all the data and information amassed on the Internet, this would form one concrete, ultimate reality (metaphysical reality). However, rather than enduring this information overload, each end-user is provisioned a filtered portion of this reality, with additional filters of perception (e.g. due to personalized preferences, goals/incentives of platform owners). 
As more of the physical world is driven by events taking place in the digital world, one could argue that a large proportion of the metaphysical reality resides in the digital reality. 
While some existing work on digital metaphysics \cite{Steinhart1998-STEDM, 10.5555/529708, Rescher1991-RESGLM} offers the proposition that this digital world (and hence the physical world) may be driven by programs (or automata) that potentially autonomously encode the rules of the physical world and may interact with other programs, 
the current state of affairs indicates that, regardless of whether programs may or may not be sentient, they do indeed have an effect on the physical world through inputs and outputs manifested by end-users. End-users are immersed in the digital world in their daily lives (users spent 161 minutes per day on mobile and desktop devices in 2018 \citep{stat_0}), are constantly tracked across platforms and apps \cite{10.1145/3201064.3201089, KollnigShubaBinnsVan} (i.e. providing inputs to the automated programs), and content generated by platforms, end-users or other third-parties are distributed and matched to end-users through recommendation systems personalized with respect to tracked user data \citep{karlgren1990algebra} (with varying levels of regulation or moderation by human / platform owners).


Having argued in Section 4.2 that the proliferation of end-user interface development/re-rendering systems like \grease{} can enable end-users to re-design and re-render interfaces according to their own preferences, objectives and goals (and potentially eliminating the goals/objectives of other network end-users, third parties, or software/platform developers), it follows that 
this enhanced end-user autonomy over their interfaces also grants end-users the ability to re-write their digital realities. By encoding rules in the screenome visualization page on what aspects of the interface to augment or occlude, the end-user is dictating which aspects of the metaphysical (ultimate) reality that they wish to observe. This, in turn, may change the cognition and end-user perception in the longer term. For example, we sample benevolent use cases in this manuscript, in particular the reduction in harmful content to users.
One could argue that the current digital reality is already personalized to end-user preferences and data, and whether marginal value of an additional layer of end-user personalization exists. The argument against this is that, though personalized to our personal data, the interfaces and their content are dictated by systems that may not take into account our goals and objectives; they may have our data, but they may not have our best interests. Conversely, by dictating our digital experiences, they have the ability to reverse the feedback loop: ideally the system should render what the user would like to see, but instead what we observe is that the system nudges the user into wanting to see what the system would like them to see. With end-user interventions, at least the end-user has an option of correcting for reverse-nudging, where the system nudges the user rather than vice versa,
and gives users the right to see what they really would like to see. 
If we continue to extend the functionality of \grease{}, we can enable the re-writing of the digital reality (thus far indicated to be connected to the physical world). For example, we could develop interventions that re-render the digital interface as time-dependent, e.g. filtering or stylizing content to be 1960s-based, to further evaluate benefits of reminiscence therapy in aged patients facing dementia \citep{Woods00}. 


The progression for maximizing end-user autonomy and personalizing digital realities can raise unexpected costs, and requires careful consideration of what regulation for these realities might look like.
While benevolent use cases exist, malevolent use cases can also arise. For example, end-user intervention development could be abused to filter holistic information about the world and further ingrain users into niche and perceptually-harmful bubbles, which is already a harmful phenomenon manifesting in social networks. 
Since it has been argued thus far that changing one's digital reality can alter one's cognition over time, the extreme personalization and decentralization of user interface design and rendering can result in adverse effects in the physical world if the personalization is unregulated. A parallel is observed in the Internet, where due to extreme decentralization of the underlying technology and protocols, pockets of the Internet exist in the dark web, and these unregulated regions manifest discussions on murder, assassinations, rape, and other heinous dangers. While discussions take place online, real-world effects leak out, and countermeasures are regularly taken by law enforcement to curb this decentralization and anonymity. 
Likewise, rather than curbing harmful content, end-users may in turn filter specifically for harmful content and increase their exposure to dangerous content that can result in harm to other individuals. For example, certain end-users may choose to block beneficial content, and choose to occlude any content relating to the topic of "women" in general, which may result in the filtering of content pertaining to women's health or systemic issues pertaining to women's rights, and over a long period of lacking exposure, may result in the individual developing sexist worldviews and tendencies. 

Hence, the next step needs to be a discussion and investigation on what alternative regulation on our digital realities need to look like. End-user personalization through browser and app extension ecosystems are being developed and deployed, though the extensions/interventions are currently still constrained by intervention development by professional developers, hence this state of criticality is still averted thus far. Currently, it is the tech companies, the social networks (driven by wisdom of crowds and virality), and governments to some extent that are collectively fighting (sometimes against each other) to regulate the platforms and software. In this manuscript, we have advocated for maximizing the goals and incentives of end-users and individual welfare, but we will need to think carefully towards social welfare as well. Regulation by tech companies may not be ideal given an intrinsic motivation towards optimizing viewership and profit, even at the cost of individual or social welfare. The effect of government regulation may vary by region and may conflict with individual or social welfare as well, such as through enhanced interface-agnostic content censorship in oppressive regimes. 

\section{Conclusion}
\label{sec:conclusion}


To enable end-user autonomy over interface design, and the generation and proliferation of a distribution of harms and interventions to analyze and reflect upon, 
we contribute the novel interface modification framework \grease{}.
End-users can reflect and annotate with their digital browsing experiences, 
and collaboratively craft interface interventions with our HITL and visual overlay mechanisms.
We hope \grease{} will enable other researchers (and eventually end-users) 
to study harms and interventions, and other interface modification use cases.

\newpage
\bibliographystyle{ACM-Reference-Format}
\bibliography{main}


\end{document}